 \documentclass[journal]{IEEEtran}
\newcommand{\qeda}{\hfill\ensuremath{\blacksquare}}

\newcommand{\m}[1]{\mbox{$#1$}}
\let\labelindent\relax
\usepackage{enumitem}

\usepackage{afterpage}

\usepackage{relsize}
\usepackage{subfigure}
\newcommand{\acap} {\hspace{.5pt}\mathlarger{\cap}\hspace{.5pt}}
\usepackage{bm}

\usepackage{setspace}
\usepackage{psfrag}
\usepackage{cite}
\usepackage{mathrsfs}
\usepackage{bbm}
\usepackage{pifont}
\usepackage{amsfonts}
\usepackage{amsmath, amsthm}
\usepackage{csquotes}
\newif\ifdraft
\draftfalse
\usepackage{tabularx}
\usepackage{listings}
\usepackage{graphicx}
\usepackage{float}
\usepackage{amssymb}
\usepackage{array}
\usepackage{fancyhdr}
\usepackage{cases}
 \usepackage{supertabular}
\usepackage{url}
\usepackage{fancyhdr}

\usepackage{color}
\usepackage{bbding}
\newcommand{\bcap} {\hspace{2pt} \mathlarger{\cap}
\hspace{2pt}}
\newcommand{\h}{it holds that }

\newcommand{\f}{it follows that }

\newcommand{\bcup} {\hspace{2pt} \mathlarger{\cup}
\hspace{2pt}}

\newcommand {\C} {{\rm I\kern-5.5pt C}}

\newcommand{\bP}[1]{{\mathbb{P}}\left[{#1}\right]}
\newcommand{\bE}[1]{{\mathbb{E}}\left[{#1}\right]}

\newcommand{\1}[1]{{\bf 1}\left[#1\right]}       % indicator 1{...}

\def\centerhack#1{\hbox to 0pt{\hss\footnotesize #1\hss}}
\def\centerhackn#1{\hbox to 0pt{\hss #1\hss}}
\def\dchack#1{\vbox to 0pt{\vss{\hbox to 0pt{\hss#1\hss}}\vss}}

\usepackage{etoolbox}

\newtheorem{fact}{Fact}
\newtheorem{lem}{Lemma}
\newtheorem{thm}{Theorem}

\newtheorem{rem}{Remark}
\newtheorem{cor}{Corollary}
\newtheorem{proposition}{Proposition}
\newtheorem*{proposition1.1}{Proposition 1.1}
\newtheorem*{proposition1.2}{Proposition 1.2}
\newtheorem*{proposition1.3}{Proposition 1.3}
\newtheorem*{proposition2.1}{Proposition 2.1}
\newtheorem*{proposition2.2}{Proposition 2.2}
\IEEEoverridecommandlockouts
\begin{document}

\title{Analyzing resilience of interest-based social networks against node and link failures}

%
%\title{\fontsize{22}{23} \selectfont Resilience and Connectivity of Secure Sensor Networks}

%\title{Component Evolution of Secure Sensor Networks with On/Off Channels}

%\title{Robust Design of Secure Sensor
%Networks}

%\title{Rigorous Designing Secure Wireless Sensor
%Networks for Connectivity}

%\title{\newttlfnt
%Connectivity in Random Key Graphs under Geometric Constraints}

%with Geometric Constraints

%\title{\newttlfnt
%Connectivity and Energy Consumption in Secure Wireless Sensor
%Networks under Key Predistribution Schemes} with Geometric
%Constraints}

\iffalse

 \author{ \IEEEauthorblockN{Jun Zhao, Osman Ya\u{g}an and Virgil Gligor}
\IEEEauthorblockA{CyLab and Dept. of ECE\\
Carnegie Mellon University \\
{\tt \{junzhao,oyagan,gligor\}@cmu.edu}}}

 Email: junzhao@alumni.cmu.edu

 \fi

%\author{Jun~Zhao\\{\tt junzhao@alumni.cmu.edu}\thanks{Jun~Zhao is a postdoctoral researcher at Arizona State University
%and Princeton University, after receiving a PhD from Carnegie Mellon University.}}

\author{Jun Zhao,~\IEEEmembership{Member,~IEEE}  % <-this % stops a space
\IEEEcompsocitemizethanks{\IEEEcompsocthanksitem J. Zhao was with Carnegie Mellon University, Pittsburgh,
PA 15213. He is now with Arizona State University, Tempe, AZ 85281. This work was supported in part by the U.S. National Science Foundation (NSF) under Grants SaTC-1618768 and CNS-1422277, and in part by Army Research Office under Grant W911NF-16-1-0448.
 Email: junzhao@alumni.cmu.edu}} %\protect\\

\maketitle

 %\thispagestyle{plain} \pagestyle{plain}

%
%\numberofauthors{1}
%\author{
%\alignauthor~\\
%       \affaddr{(The full version)\\}}
%

%\numberofauthors{1}
%\author{
%\alignauthor ~\\}

%\numberofauthors{1}
%\author{
%\alignauthor ~\\
%       \affaddr{Paper ID: 4}}

%\numberofauthors{3}
%\author{
%\alignauthor
%Paper ID\\
%       \affaddr{Authors to be anonymized}\\
%       \affaddr{(Zhao, Ya\u{g}an and Gligor)}
%       }
%\numberofauthors{3}
%\author{
%\alignauthor
%Jun Zhao\\
%       \affaddr{CyLab and Dept.
%of ECE}\\
%       \affaddr{Carnegie Mellon University}\\
%       \affaddr{Pittsburgh, PA 15213}\\
%       \email{junzhao@cmu.edu}
%\alignauthor {\aufnt Osman Ya\u{g}an}\\
%       \affaddr{CyLab and Dept.
%of ECE}\\
%       \affaddr{Carnegie Mellon University}\\
%       \affaddr{Moffett Field, CA 94035}\\
%       \email{oyagan@ece.cmu.edu}
%\alignauthor Virgil Gligor\\
%       \affaddr{CyLab and Dept.
%of ECE}\\
%       \affaddr{Carnegie Mellon University}\\
%       \affaddr{Pittsburgh, PA 15213}\\
%       \email{gligor@cmu.edu}}
\maketitle
%\thispagestyle{plain} \pagestyle{plain}

%\vspace{80pt}

 \begin{abstract}

 In typical online social networks, users are linked by symmetric friend relations and can define circles of friends based on shared interests. In this paper, we look at social networks where users form links subject to both friendships and shared interests. Our goal is to understand resilience of these networks in terms of connectivity when both nodes and links are allowed to fail. We derive a zero-one law as well as the asymptotically exact probability result for connectivity under both node and link failures. The results answer the question of how to set the network
parameters such that reliable message dissemination can be achieved. We formally prove the results and confirm the results via experiments as well.

\end{abstract}

%a protein similarity network.


\section{Introduction}

\begin{table*}[!t]
\centering
\caption{Different constraints and their induced graph  topologies.}
\label{my-label}
\begin{tabular}{|l|l|}
\hline
Constraints                                                                                                                                                                                                                                                          & Induced graph                            \\ \hline
\begin{tabular}[c]{@{}l@{}}common-interest relations:\\ each user \emph{independently} selects $K_n$ objects \emph{uniformly at random} from the same pool of $P_n$ objects;\\ two users establish a common-interest relation if and only if they share at least $d$ object(s).\end{tabular} & $G_d(n, K_n,P_n)$                        \\ \hline
\begin{tabular}[c]{@{}l@{}}social friendships:\\ two users are friends of each other with probability $f_n$.\end{tabular}                                                                                                                                            & $G(n,{f_n\iffalse_{on}\fi})$             \\ \hline
\begin{tabular}[c]{@{}l@{}}link failures:\\ the link between two users fails with probability $1-g_n$, and remains with probability $g_n$.\end{tabular}                                                                                                                                                 & $G(n,{g_n\iffalse_{on}\fi})$          \\ \hline
\begin{tabular}[c]{@{}l@{}}common-interest relations \& social friendships \& link failures:\\ our studied system model in consideration of these three types of constriants.\end{tabular}                                                                            & $\mathbb{G}_d(n, K_n, P_n,{f_n},g_n)$ \\ \hline
\end{tabular}
\end{table*}

Online social networks interconnect users by symmetric \emph{friend}
relations \cite{kumar2010structure} and allow them to define \emph{circles of friends} (viz.,
\textit{Google+}) \cite{yang2012circle,qian2014personalized,mcauley2012learning }. We view a user's circle of friends as the group of
friends who share a \emph{common interest}. A basic common interest
between two friends can be represented by their selection of a
number of common objects from a large pool of available objects. For
example, two friends may pick the same set of songs to listen to from
\emph{Spotify}'s pool, or the same videos to watch from \textit{Youtube}'s pool, or
the same games to play from \emph{Playfire}.
%Of course, a user can belong to multiple circles of
%friends defined around the same pool of common-interest objects.
 Identifying friends with common interests in a social network
enables the implementation of large-scale, distributed
publish-subscribe services which support dissemination of
special-interest messages among the users \cite{kato2011scalable,kayastha2011applications}. Such services allow
publisher nodes to post interest-specific news, recommendations,
warnings, or announcements to subscriber nodes in a wide variety of
applications ranging from on-line behavioral advertising (e.g., the
message may contain an advertisement targeted to a common-interest
group) to social science (e.g., the message may contain a survey
request or result directed to a special-interest group).

Links and nodes in social networks can fail due to voluntary deletion \emph{Facebook} or adversarial attacks \cite{linkedin}. In this paper, we aim to understand how resilient  interest-based social networks are against both link and node failures in terms of connectivity. To answer this question, below in this section, we first detail our model for interest-based social networks, then discuss link and node failures, and finally formalize the study of resilience as a theoretical problem.

\textbf{Modeling interest-based social networks.}
Consider an undirected social network of $n$ users (all networks/graphs in the paper are undirected). The common-interest {relation} in the social
  network induces a graph $\mathcal{G}$, where each of the $n$ users
represents a node in $\mathcal{G}$ and two nodes are {connected by an
edge} if and only if the users they represent are common-interest
friends. The relevance of
the connectivity properties of $\mathcal{G}$
in the context of large-scale, distributed publish-subscribe services
can be seen as follows. Each
publisher as well as each subscriber represents a
node in $\mathcal{G}$. When publisher $v_a$ posts an interest-specific
message $\verb|msg|$, each node $v_b$ in $v_a$'s circle of
common-interest friends receives \verb|msg| and posts \verb|msg| to
its own circle of common-interest friends, unless \verb|msg| has
already been posted there recently. This process continues
iteratively. \iffalse Therefore, a subscriber receives \verb|msg|
from $v_a$ either directly or indirectly via the help of other
subscribers acting as relays. \fi
 Obviously, the global dissemination of message \verb|msg| can be
achieved if and only if there exists a path between $v_a$ and each
subscriber among the other $(n-1)$ nodes of $\mathcal{G}$, which happens
if $\mathcal{G}$ is connected, since connectivity means that any two nodes can find at least one path in between. Furthermore, even if at most $(k-1)$ users
leave the network, $k$-connectivity of $\mathcal{G}$ assures the
availability of message-dissemination path(s) between any two
remaining nodes, since $k$-connectivity is defined such
that the network remains connected despite the removal of any $(k-1)$
nodes \cite{Citeulike:505396} (removing nodes also remove their associated edges).

A possible way to construct the graph $\mathcal{G}$ on $n$ users is as follows.
Suppose that there exists an \textit{object pool}
$\mathcal {P}_n$ consisting of $P_n$ objects and that each user picks
exactly $K_n$ distinct objects uniformly and independently from the
object pool;
%\footnote{ Uniform and independent object selection by
%users yields a small lower bound on $K_n$ and still assures
%connectivity of the induced graph $\mathcal{G}$ (defined
%afterwards)~\cite{r4}.}
i.e., each user has an \textit{object ring}
consisting of $K_n$ objects (we index the parameters by $n$ to study the scaling behavior when $n$ gets large). Two {\em friends} are said to have a common-interest
relation if they have at least $d$ common objects in their object
rings. The topology induced by common-interest
relations, denoted by $G_d(n, K_n,P_n)$, is known in the literature as a uniform $d$-intersection graph  \cite{bloznelis2013,zhao2015resilience,Rybarczyk,ICASSP17-design,bradonjic2010component,ICASSP17-social}. In order to model the friendship network, we use an Erd\H{o}s-R\'enyi graph model
 as in a few
prior studies  \cite{citeulike:691419,etesami2016complexity,korolov2015actions,brummitt2015jigsaw}. Although this model is simple, we will show that when it is coupled with common-interest relations, the induced analysis   becomes quite involved. A future direction is to consider more complex models. Under the model of an
Erd\H{o}s-R\'enyi graph to represent the social network,
any two users in the network are friends with each other with probability
$f_n$ independently from all other users.
As a result, the graph $\mathcal{G}$ to model the common-interest-based subgraph of the social network becomes the intersection of an
Erd\H{o}s-R\'enyi graph $G(n,{f_n\iffalse_{on}\fi})$ and a uniform $d$-intersection graph $G_d(n, K_n,P_n)$, where the intersection of two graphs $G_1$ and $G_2$ defined on the
same node set has the following meaning:
 two nodes have an edge in between in
$G_1\bcap G_2$ if and only if these two nodes have an edge in $G_1$
and also have an edge in $G_2$. We denote the above graph $\mathcal{G}$ by $\mathcal{G}(n, K_n, P_n,
{f_n},d)$ to elipticity express its parameters; i.e.,
\begin{align}
\mathcal{G}(n, K_n, P_n,
{f_n},d) : = G_d(n, K_n,P_n) \bcap G(n,{f_n\iffalse_{on}\fi}). \label{Gintersect1}
\end{align}

\textbf{Interest-based social networks under link failure.}
To consider the resilience of the common-interest social network $\mathcal{G}(n, K_n, P_n,
{f_n},d)$ against link failure, we consider a simple model where each link fails independently with probability $1-g_n$; i.e., under link failure, each link is  preserved with probability $g_n$. Link failure in   social networks may result from adversarial attacks \cite{shrivastava2008mining,li2012measuring,sagduyu2015navigating,conti2013virtual}. Then the graph model for the common-interest social network under link failure is obtained by further superimposing an
Erd\H{o}s-R\'enyi graph $G(n,{g_n\iffalse_{on}\fi})$ over $\mathcal{G}(n, K_n, P_n,
{f_n},d)$. Letting $\mathbb{G}_d(n, K_n, P_n,
{f_n},g_n)$ denote the induced graph model, we obtain
\begin{align}
\mathbb{G}_d(n, K_n, P_n,
{f_n},g_n) : = \mathcal{G}(n, K_n, P_n,
{f_n},d) \bcap G(n,{g_n\iffalse_{on}\fi}). \label{Gintersect2}
\end{align}
Substituting (\ref{Gintersect1}) into (\ref{Gintersect2}), we further have
\begin{align}
&\mathbb{G}_d(n, K_n, P_n,
{f_n},g_n) \nonumber \\  & : = G_d(n, K_n,P_n) \bcap G(n,{f_n\iffalse_{on}\fi})\bcap G(n,{g_n\iffalse_{on}\fi}). \label{Gintersect3}
\end{align}
Table I summarizes the graph notation.

%
%$$\mathcal{G}(n, K_n, P_n,
%{f_n},d) : = G_d(n, K_n,P_n) \bcap G(n,{f_n\iffalse_{on}\fi}).$$
%
%$$\mathbb{G}_d(n, K_n, P_n,
%{f_n},g_n) : = \mathcal{G}(n, K_n, P_n,
%{f_n},d) \bcap G(n,{g_n\iffalse_{on}\fi}).$$
%
%$$\mathbb{G}_d(n, K_n, P_n,
%{f_n},g_n) : = \mathcal{G}(n, K_n, P_n,
%{f_n},d) \bcap G(n,{g_n\iffalse_{on}\fi}).$$

%, and simplify the notation by writing $\mathcal{G}$ (we use $\mathcal{G}$ instead of $\mathcal{G}$ to reflect its dependence on $d$).

\textbf{Interest-based social networks under both link and node failures.} As explained above, an interest-based social network under link failure is modeled by graph $\mathbb{G}_d(n, K_n, P_n,
{f_n},g_n) $. To study node failure, our goal is to ensure that given some $m<n$, graph $\mathbb{G}_d(n, K_n, P_n,{f_n},g_n)$ remains connected even after an arbitrary set of $m$ nodes fail, where connectivity means that any two nodes can find a path in between. In other words, our goal is to ensure that for an interest-based social network $\mathcal{G}(n, K_n, P_n,
{f_n},d)$, after each link is independently deleted with probability $1-g_n$, and then after an arbitrary set of $m$ nodes fail, the remaining graph is still connected. Note that deleting a link does not remove its endpoints from the network, but removing a node will delete both the node and all of its links from the network.

\textbf{Research problem: How resilient are interest-based social networks against both link and node failures in terms of connectivity?} Our goal is to understand how resilient are interest-based social networks against both link and node failures in terms of connectivity. An interest-based social network under link failure is modeled by graph $\mathbb{G}_d(n, K_n, P_n,
{f_n},g_n) $, as explained above.
We will study connectivity behavior of graph $\mathbb{G}_d(n, K_n, P_n,
{f_n},g_n)$ when an arbitrary set of $m$ nodes can fail. Under node failure, we remove the failed nodes and their associated links from the graph. We will derive a zero-one law and the asymptotically exact probability result for $\mathbb{G}_d(n, K_n, P_n,
{f_n},g_n)$ in the presence of node failure, where the zero-law (resp., one-law) shows that the remaining
graph is  disconnected (resp., connected) asymptotically. Our results enable
us to answer the two key questions for the design of a large-scale,
reliable publish-subscribe service: (1) what values should the
parameters  $n$, $K_n$, $P_n$, $f_n$ and $d$ take in order to achieve connectivity
between publisher and subscriber nodes in the common-interest graph
$\mathcal{G}$; and (2) how can reliable message dissemination be achieved
when links and nodes are both allowed to fail. These failures could happen as a
result of discretionary user action (e.g., a node may decide not to
forward a particular message, or all messages, of a particular
publisher); or voluntary account deletion (e.g., \emph{Facebook}
account deletions are not uncommon events~\cite{quitFacebook}); or
involuntary account deletion caused by adversary attacks (e.g.,
Agarwalla \cite{linkedin} shows that clickjacking vulnerability
found in Linkedin results in involuntary account deletion).

\textbf{Roadmap.}
We organize the rest of the paper as follows. We detail the analytical results as Theorem \ref{thm:OneLaw+NodeIsolation-resi} in Section
\ref{sec:res}. Subsequently, we provide   experiments in
Section \ref{sec:expe} to confirm our analytical results.  Section \ref{related} surveys related work. After introducing some preliminaries in  Section  \ref{sec:SystemModel}, we discuss in  Section  \ref{prf_idea_thm:OneLaw+NodeIsolation} the   ideas to establish  Theorem \ref{thm:OneLaw+NodeIsolation-resi}. We conclude the paper in Section \ref{sec:Conclusion}. Many technical details are provided in the Appendix for clarity.

\section{The Results} \label{sec:res}

We present and discuss our results in this section.
%$\mathbb{N}_0 $ stands for the set of all positive integers;
%$\mathbb{R}$ is the set of all real numbers;
The
natural logarithm function is given by $\ln$. All limits are understood with $n
\to
  \infty$.  %The term ``for all $n$
%sufficiently large'' means ``for any $n \geq N$, where $N \in
%\mathbb{N}_0$ is selected appropriately''.
 We use the standard
asymptotic notation $o(\cdot), O(\cdot), \Omega(\cdot), \omega(\cdot),
\Theta(\cdot), \sim$; see \cite[Page 2-Footnote 1]{ZhaoYaganGligor}. Throughout the paper, $m$ and $d$ are positive constant integers so they do
not scale with $n$.  The notation $\mathbb{P}[\mathcal {E}]$ denotes the
probability that an event $\mathcal {E}$ happens.

Theorem \ref{thm:OneLaw+NodeIsolation-resi} below presents a zero-one law as well as the asymptotically exact probability result for connectivity in a  graph $\mathbb{G}_d(n, K_n, P_n,{f_n},g_n)$ under node failure. Note that $g_n$ (more precisely, its complement $1-g_n$) in $\mathbb{G}_d(n, K_n, P_n,{f_n},g_n)$ already encodes link failure. The zero-law means that the probability of connectivity asymptotically converges to $0$ under some conditions and the one-law means that the probability of connectivity asymptotically converges to $1$ under some other conditions.

\begin{thm}  \label{thm:OneLaw+NodeIsolation-resi}

For a graph $\mathbb{G}_d(n, K_n, P_n,{f_n},g_n)$ which models a common-interest social network under \textbf{link failure}, with a sequence $\alpha_n$ defined by
\begin{align}
{f_n} \cdot g_n \cdot   \sum_{u=d}^{K_n}
\frac{\binom{K_n}{u}\binom{P_n-K_n}{K_n-u}}{\binom{P_n}{K_n}} & = \frac{\ln  n +m\ln \ln n   +
 {\alpha_n}}{n},   \label{eq:scalinglaw-resi}
\end{align}
  it holds under $ K_n =
\Omega(n^{\epsilon})$ for a positive constant $\epsilon$, $ \frac{{K_n}^2}{P_n}  =
 o\left( \frac{1}{\ln n} \right)$,  and $ \frac{K_n}{P_n} = o\left( \frac{1}{n\ln n} \right)$  that
\iffalse
\begin{subnumcases}
{ \hspace{-7pt}  \lim_{n \rightarrow \infty }\hspace{-2pt} \mathbb{P}\hspace{-1.5pt}\bigg[
\hspace{-4.9pt}\begin{array}{c}
\mathbb{G}_d(n, K_n, P_n,{f_n},g_n) \\
\text{is connected even after an arbitrary set of $m$ \bm{nodes fail}.}
\end{array}\hspace{-4.9pt}
\bigg] \hspace{-3.7pt}=\hspace{-3.7pt}}  \hspace{-5pt}0,\quad\hspace{-9pt}\text{if  }\hspace{-5pt}\lim_{n \to \infty}\hspace{-1.5pt}{\alpha_n}  \hspace{-2.5pt} =  \hspace{-2.5pt} - \infty, \label{thm-con-eq-0-resi} \\
\hspace{-5pt}1,\quad\hspace{-9pt}\text{if  }\hspace{-5pt}\lim_{n \to \infty}\hspace{-1.5pt}{\alpha_n}    \hspace{-2.5pt}=  \hspace{-2.5pt}  \infty, \label{thm-con-eq-1-resi}
\end{subnumcases}
\fi
\begin{align}
& \hspace{-21pt} \lim_{n \rightarrow \infty } \hspace{-1pt} \mathbb{P}\bigg[
\hspace{-2.9pt}\begin{array}{c}
\mathbb{G}_d(n, K_n, P_n,{f_n},g_n) \mbox{ remains connected} \\
\mbox{even after an arbitrary set of $m$ \textbf{nodes fail}.}
\end{array}\hspace{-2.9pt}
\bigg]  \nonumber
% \\  & \hspace{-27pt}
% =   e^{- \frac{e^{-\lim_{n \to \infty}{\alpha_n}}}{(k-1)!}}  \label{thm-con-eq-compact}
\\ &  \hspace{-21pt}  =  e^{- \frac{e^{-\lim_{n \to \infty} \alpha_{_n}}}{m!}} \label{thm-con-eq-resi-compact}
\end{align}
\vspace{-5pt}
\begin{subnumcases}
{ \hspace{-47pt}=\hspace{-2pt}} \hspace{-3pt}e^{- \frac{e^{-\alpha ^*}}{m!}},
 &\hspace{-11.5pt}\text{ if $\lim\limits_{n \to \infty}{\alpha_n}
=\alpha ^* \in (-\infty, \infty)$,} \label{thm-con-eq-e-resi} \\  \hspace{-3pt}1, &\hspace{-11.5pt}\text{ if $\lim\limits_{n \to \infty}{\alpha_n}
=\infty$,} \label{thm-con-eq-1-resi}  \\ \hspace{-3pt} 0, &\hspace{-11.5pt}\text{ if $\lim\limits_{n \to \infty}{\alpha_n}
=-\infty$}. \label{thm-con-eq-0-resi}  \label{thm-con-eq-e-resi}
\end{subnumcases}

\end{thm}

\textbf{Interpreting Theorem \ref{thm:OneLaw+NodeIsolation-resi}.} For the property that $\mathbb{G}_d(n, K_n, P_n,{f_n},g_n)$ remains connected even after an arbitrary set of $m$ {nodes fail}, (\ref{thm-con-eq-resi-compact}) of Theorem \ref{thm:OneLaw+NodeIsolation-resi} presents the asymptotically exact
probability, while (\ref{thm-con-eq-1-resi}) and (\ref{thm-con-eq-0-resi}) of Theorem \ref{thm:OneLaw+NodeIsolation-resi} together constitute a zero--one law, where a zero--one law means that the probability of a graph having a certain property asymptotically converges to $0$ under some conditions and to $1$ under some other conditions.  The result (\ref{thm-con-eq-resi-compact}) compactly summarize (\ref{thm-con-eq-e-resi})--(\ref{thm-con-eq-0-resi}).

As will be clear in Section \ref{sec:SystemModel}, the left hand side of (\ref{eq:scalinglaw-resi}) equals the edge probability $\mathbb{G}_d(n, K_n, P_n,{f_n},g_n)$, where the edge probability is the probability that two nodes have an edge in between (note that for any pair of nodes in $\mathbb{G}_d(n, K_n, P_n,{f_n},g_n)$, the probability of having an edge in between is the same).

 Theorem \ref{thm:OneLaw+NodeIsolation-resi} shows that a critical scaling
  for connectivity in graph $\mathbb{G}_d(n, K_n, P_n,{f_n},g_n)$ under the failure of $m$ nodes is that the left hand side of (\ref{eq:scalinglaw-resi}) equals  $\frac{\ln  n +m\ln \ln n  }{n}$. When $\mathbb{G}_d(n, K_n, P_n,{f_n},g_n)$ is connected even after an arbitrary set of $m$ nodes fail, we can equivalently say that $\mathbb{G}_d(n, K_n, P_n,{f_n},g_n)$ is $(m+1)$-connected (i.e., $k$-connected for $k=m+1$), from the definition of $(m+1)$-connectivity~\cite{Citeulike:505396}.

We discuss the practicality of the conditions   in Theorem \ref{thm:OneLaw+NodeIsolation-resi}: $ K_n =
\Omega(n^{\epsilon})$ for a positive constant $\epsilon$, $ \frac{{K_n}^2}{P_n}  =
 o\left( \frac{1}{\ln n} \right)$,  and $ \frac{K_n}{P_n} = o\left( \frac{1}{n\ln n} \right)$.
All conditions are enforced here
merely for technical reasons, but we   explain that they hold  in realistic
social network applications. It is expected \cite{zhang2015social}
that the object pool size $P_n$ will be much larger than  the number $n$
of participating users, which will be further larger than the number $K_n$ of objects associated with each user. We note that the condition on $K_n$ in Theorem \ref{thm:exact_qcomposite-kcon} (i.e., $ K_n =
\Omega(n^{\epsilon})$ is less appealing but is not much a problem because $\epsilon$ can be arbitrarily small. Also, this condition can be improved (i.e., weakened) to the more practical $ K_n =
\omega(\ln n)$, by trading-off the granularity of the $k$-connectivity results; we provide the details in Appendix I of the full version \cite{fullpdfaaaia}.

  In  Section  \ref{prf_idea_thm:OneLaw+NodeIsolation}, we discuss  the   ideas to establish  Theorem \ref{thm:OneLaw+NodeIsolation-resi}, after presenting experiments in Section  \ref{sec:expe} and  preliminaries in  Section  \ref{sec:SystemModel}.

% in the full version \cite{fullpdfaaai} due to space limitation.

\section{Experimental Results}
\label{sec:expe}

To confirm our analytical results in Theorem \ref{thm:OneLaw+NodeIsolation-resi} of Section~\ref{sec:res}, we perform experiments on real-world social networks as well as synthetic networks to plot a few figures.

We first present experimental results on real social networks.
We analyze data crawled from www.Anobii.com, a website for book lovers (primarily popular in Italy). In Anobii, every user can build personal digital library by picking book titles (that the user is interested in reading)
from a large database of books. Every user can also find out which other users are reading the same book with s/he; i.e., users with common interests can be identified. In addition, Anobii suggests to establish friendships with people that a user already knows
on Facebook; i.e., users on Anobii can also have friendships. Since Anobii has two different kinds of ties, namely the common-interest relations and friendships, Anobii provides an exemplary case for our study of interest-based social networks.

In Anobii, we consider books published after 2010 to have the experiments tractable. This gives us about a pool of \mbox{$9 \times 10^4$} books. We crawled Anobii in December 2016 and obtained a dataset of about $6 \times 10^4$ users. To get an interest-based social network from the data, we enforce that two users establish a link in between if and only if they read at least $2$ common books and are friends with each other. On average, a user's library has 10 books published after 2010,  while a user has about 300 social friends on average. We perform experiments on this network to confirm the theoretical results of $\mathbb{G}_d(n, K, P,
{f},g) $ with $n=6\times 10^4$ (since there are $6 \times 10^4$ users), $K=10$ (since a user has 10 books on average), $P = 9\times 10^4$ (since there are $9 \times 10^4$ books in total), $d = 2$ (since a common-interest relation requires the sharing of at least $2$ common books), the friendship probability $f=0.005$ (explained below)  and the link-active probablity $g$ to be set later. To obtain $f=0.005$, noting that a user in an Erd\H{o}s-R\'enyi graph $G(n,{f_n})$ modeling the friendship network has $(n-1)f_n$ friends on average, we set $(n-1)f_n$ as $300$ since a user has about 300 social friends on average from the data; this gives $f_n = \frac{300}{n-1}= \frac{300}{6\times 10^4-1}\approx 0.005$.

\afterpage{
\begin{figure}[!t]
 \centerline{\includegraphics[width=0.45\textwidth]{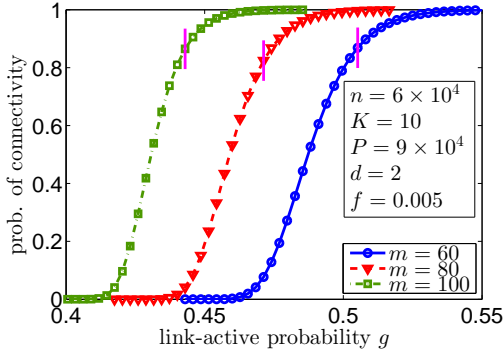}} \vspace{-10pt}\caption{Experiments on a real Anobii network with different link-active probabilities. In the Anobii network, two users establish a link in between if and only if they read at least $2$ common books and are friends with each other. After the random process of deleting each link with probability $1-g$, the Anobii network under link failure provides an understanding of an interest-based social graph $\mathbb{G}_d(n, K, P,
{f},g) $ with $n=6\times 10^4$, $K=10$, $P = 9\times 10^4$, $d = 2$, and $f=0.005$. This plot presents the probability
that the Anobii network under random link failure is connected even after an arbitrary set of $m$ {nodes fail}, as a function of $g$ for $m=60,80,100$.
}
 \label{figure1}
\end{figure}

\begin{figure}[!t]
 \centerline{\includegraphics[width=0.45\textwidth]{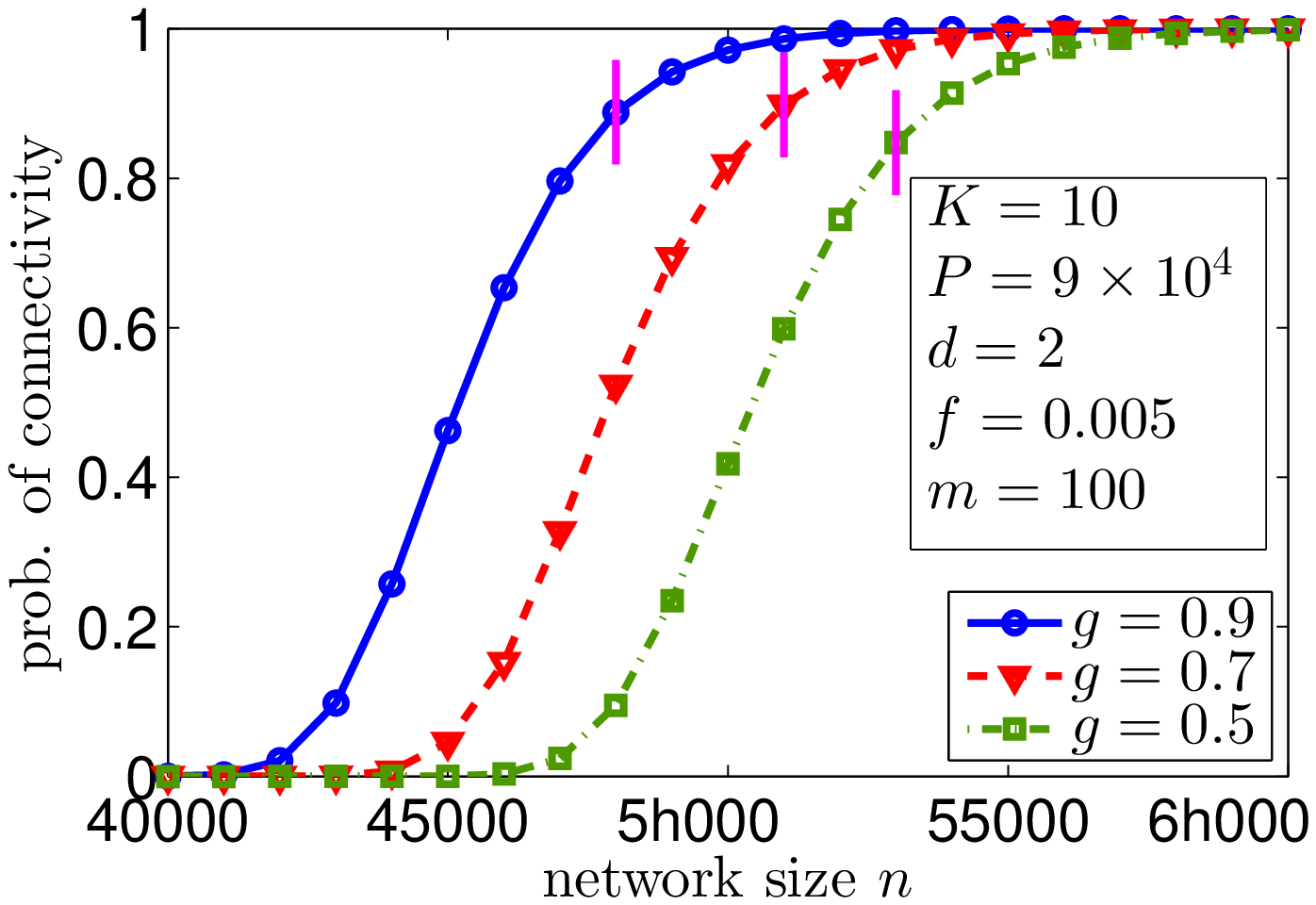}} \vspace{-10pt}\caption{Experiments on different scales of real Anobii networks. In each Anobii network, two users establish a link in between if and only if they read at least $2$ common books and are friends with each other. After the random process of deleting each link with probability $1-g$, each Anobii network under link failure provides an understanding of an interest-based social graph $\mathbb{G}_d(n, K, P,
{f},g) $ with $K=10$, $P = 9\times 10^4$, $d = 2$, and $f=0.005$. This plot presents the probability
that each Anobii network under random link failure is connected even after an arbitrary set of $m$ {nodes fail}, as a function of $n$ for $g=0.9,0.7,0.5$.
}
 \label{figure2}
\end{figure}

\begin{figure}[!t]
 \centerline{\includegraphics[width=0.45\textwidth]{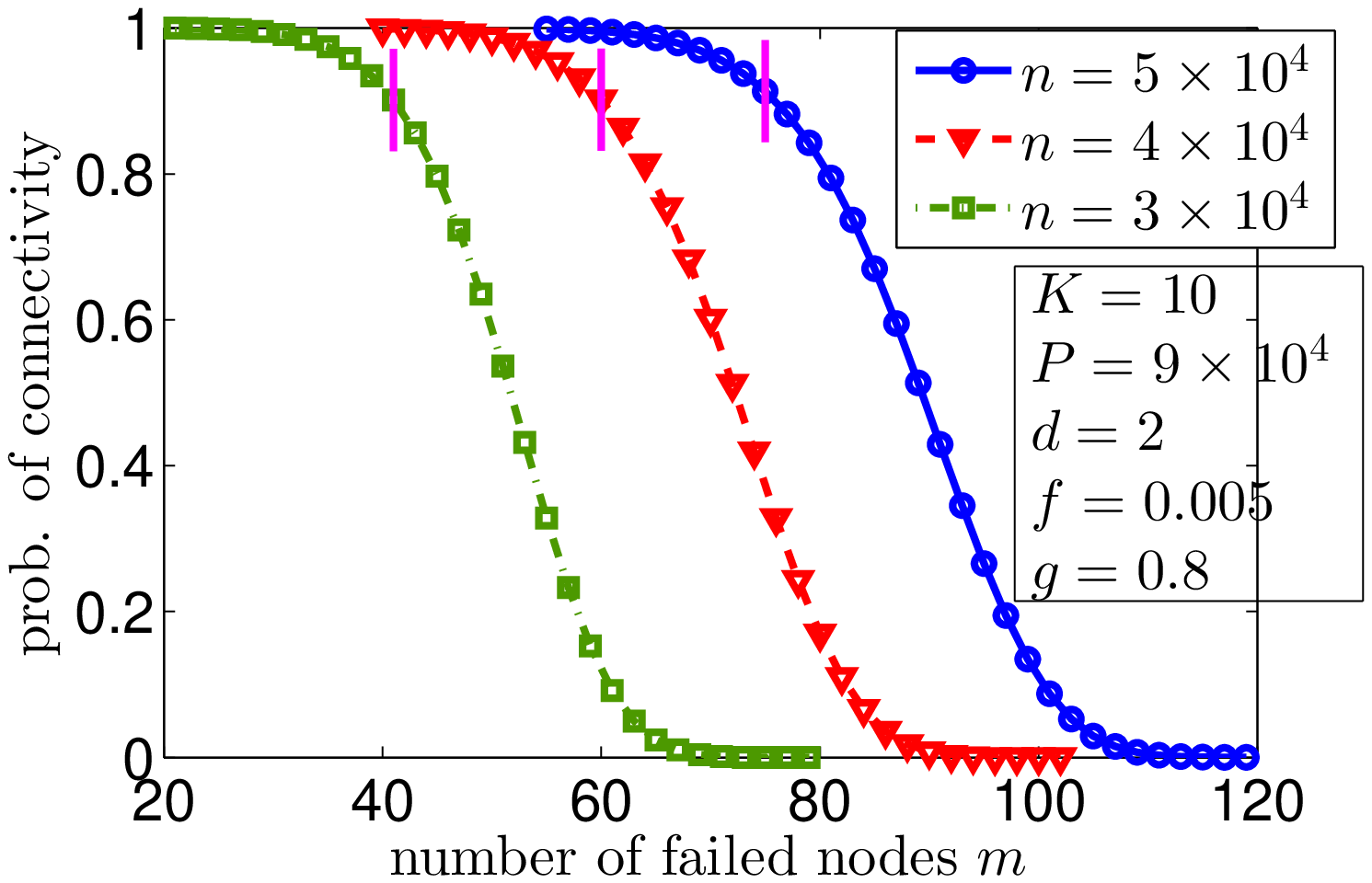}} \vspace{-10pt}\caption{Experiments on different scales of real Anobii networks. In each Anobii network, two users establish a link in between if and only if they read at least $2$ common books and are friends with each other. After the random process of deleting each link with probability $1-g$, each Anobii network under link failure provides an understanding of an interest-based social graph $\mathbb{G}_d(n, K, P,
{f},g) $ with $K=10$, $P = 9\times 10^4$, $d = 2$,  $f=0.005$ and $g=0.8$. This plot presents the probability
that each Anobii network under random link failure is connected even after an arbitrary set of $m$ {nodes fail}, as a function of $m$ for $n=5\times 10^4, 4\times 10^4, 3\times 10^4$.
}
 \label{figure3}
\end{figure}

\begin{figure}[!t]
 \centerline{\includegraphics[width=0.5\textwidth]{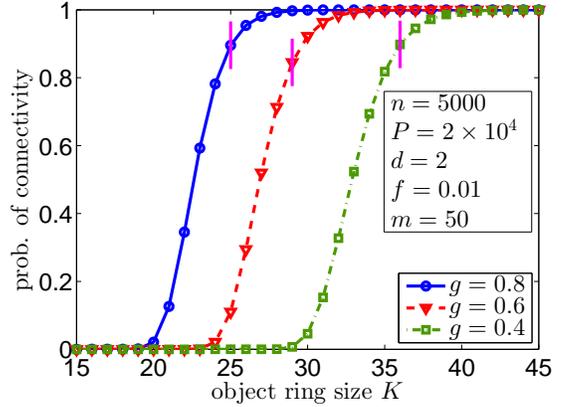}} \vspace{-10pt}\caption{Experiments on synthetic networks. Each synthetic network is sampled from the interest-based social graph model $\mathbb{G}_d(n, K, P,
{f},g) $ with $n=5000$, $P = 2\times 10^4$, $d = 2$, and $f=0.01$, $m=50$. This plot presents the probability
that $\mathbb{G}_d(n, K, P,
{f},g) $ is connected even after an arbitrary set of $m$ {nodes fail}, as a function of $K$ for $g=0.8,0.6,0.4$. Each probability is
obtained by averaging over $1000$ samples.
}
 \label{figure4}
\end{figure}

\begin{figure}[!t]
 \centerline{\includegraphics[width=0.5\textwidth]{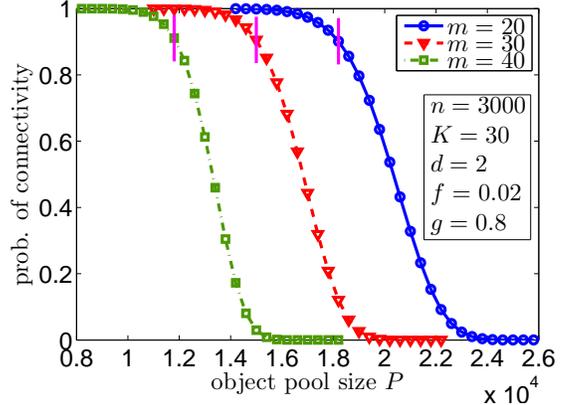}} \vspace{-10pt}\caption{Experiments on synthetic networks. Each synthetic network is sampled from the interest-based social graph model $\mathbb{G}_d(n, K, P,
{f},g) $ with $n=3000$, $K=30$, $d = 2$, $f=0.02$,  and $g=0.8$. This plot presents the probability
that $\mathbb{G}_d(n, K, P,
{f},g) $ is connected even after an arbitrary set of $m$ {nodes fail}, as a function of $P$ for $m=20,30,40$. Each probability is
obtained by averaging over $1000$ samples.
}
 \label{figure5}
\end{figure}

\begin{figure}[!t]
 \centerline{\includegraphics[width=0.5\textwidth]{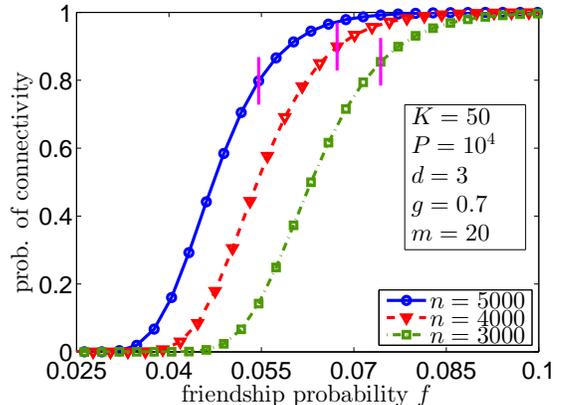}} \vspace{-10pt}\caption{Experiments on synthetic networks. Each synthetic network is sampled from the interest-based social graph model $\mathbb{G}_d(n, K, P,
{f},g) $ with $K=50$, $P = 10^4$, $d = 3$, and $g=0.7$, $m=20$. This plot presents the probability
that $\mathbb{G}_d(n, K, P,
{f},g) $ is connected even after an arbitrary set of $m$ {nodes fail}, as a function of $f$ for $n=5000,4000,3000$. Each probability is
obtained by averaging over $1000$ samples.
}
 \label{figure6}
\end{figure}

\clearpage
}

In Figure \ref{figure1}, we perform experiments on the Anobii network while varying the link-active probablity $g$; i.e., we delete each link in the Anobii network independently with probability $1-g$ to see the impact of link failure. The  Anobii network under link failure provides an understanding of an interest-based social graph $\mathbb{G}_d(n, K, P,
{f},g) $. We follow this random process of link deletion to generate $1000$ independent samples, record the count that the obtained network is connected even after an arbitrary set of $m$ nodes fail, and then divide the count by $1000$ to obtain the empirical probability of the Anobii network under link failure being connected even after an arbitrary set of $m$ {nodes fail}. We plot these probabilities in Figure~\ref{figure1}, which provides an understanding for the probability of $\mathbb{G}_d(n, K, P,
{f},g) $ remaining connected even after an arbitrary set of $m$ {nodes fail}. We clearly observe the transitional behavior in Figure \ref{figure1}. In addition, each vertical line in Figure \ref{figure1} presents the
\emph{critical} parameter $g^*$ computed from ${f}\cdot g^* \cdot  \sum_{u=d}^{K}
\frac{\binom{K}{u}\binom{P-K}{K-u}}{\binom{P}{K}}=
 \frac{\ln n +  m\ln \ln n }{n}$ based on (\ref{eq:scalinglaw-resi}).
 We also see  that the transition point of each curve is around the critical parameter illustrated by the vertical line.  Hence, Figure \ref{figure1} is   consistent with our Theorem~\ref{thm:OneLaw+NodeIsolation-resi}.

In Figure \ref{figure2}, we use the $6 \times 10^4$-user Anobii network to generate random subgraphs of different sizes and thus vary the network size $n$. We refer to these random subgraphs of different sizes as different scales of Anobii networks.   The  Anobii network under link failure provides an understanding of an interest-based social graph $\mathbb{G}_d(n, K, P,
{f},g) $. For each parameter set, we use the random process of sampling subgraphs to generate $100$ independent copies, and then apply the random process of link deletion on each copy to generate $1000$ independent samples, so we consider $100 \times 1000$ samples in total. Afterwards, we record the count that the obtained network is connected even after an arbitrary set of $m$ nodes fail, and then divide the count by $100 \times 1000$ to obtain the empirical probability of the Anobii network under link failure being connected even after an arbitrary set of $m$ {nodes fail}. We plot these probabilities in Figure \ref{figure2}, which provides an understanding for the probability of $\mathbb{G}_d(n, K, P,
{f},g) $ remaining connected even after an arbitrary set of $m$ {nodes fail}. We clearly observe the transitional behavior in Figure \ref{figure2}. In addition, each vertical line in Figure \ref{figure2} presents the
\emph{critical} (i.e., minimal) parameter $n^*$ satisfying ${f}\cdot g \cdot  \sum_{u=d}^{K}
\frac{\binom{K}{u}\binom{P-K}{K-u}}{\binom{P}{K}} \geq
 \frac{\ln n^* +  m\ln \ln n^* }{n^*}$ based on (\ref{eq:scalinglaw-resi}).
 We also see  that the transition point of each curve is around the critical parameter illustrated by the vertical line.  Hence, Figure \ref{figure2} is in accordance with our Theorem \ref{thm:OneLaw+NodeIsolation-resi}.

The explanation of Figure \ref{figure3} is similar to that of Figure~\ref{figure2}. The difference is we vary $m$ given different $n$ in Figure~\ref{figure3}, while we vary $n$ given different $g$ in Figure \ref{figure2}. We clearly observe the transitional behavior in Figure \ref{figure3}. In addition, each vertical line in Figure \ref{figure3} presents the
\emph{critical} (i.e., maximal) parameter $m^*$ satisfying ${f}\cdot g \cdot  \sum_{u=d}^{K}
\frac{\binom{K}{u}\binom{P-K}{K-u}}{\binom{P}{K}} \geq
 \frac{\ln n +  m^*\ln \ln n}{n}$ based on (\ref{eq:scalinglaw-resi}).
 We also see  that the transition point of each curve is around the critical parameter illustrated by the vertical line.  Hence, Figure \ref{figure3} is in agreement with our Theorem \ref{thm:OneLaw+NodeIsolation-resi}.

 We have presented experiments on real-world Anobii networks. We now discuss experiments on synthetic networks where we further vary the object ring size $K$, the object pool size $P$, and the friendship probability $f$.

In Figures \ref{figure4}--\ref{figure6}, we perform experiments on synthetic networks. Each synthetic network is independently sampled from the interest-based social graph model $\mathbb{G}_d(n, K, P,
{f},g) $. Figures \ref{figure4}--\ref{figure6} present the probability
that $\mathbb{G}_d(n, K, P,
{f},g) $ is connected even after an arbitrary set of $m$ {nodes fail}. We vary $K$ given different $g$ in Figure \ref{figure4}, vary $P$ given different $m$ in Figure \ref{figure5}, and vary $f$ given different $n$ in Figure \ref{figure6}. Each probability is
obtained by averaging over $1000$ samples. In Figures \ref{figure4}--\ref{figure6}, we clearly observe the transitional behavior.  In addition, each vertical line in Figure \ref{figure4} presents the
\emph{critical} (i.e., minimal) parameter $K^*$ satisfying ${f}\cdot g \cdot  \sum_{u=d}^{K^*}
\frac{\binom{K^*}{u}\binom{P-K^*}{K^*-u}}{\binom{P}{K^*}} \geq
 \frac{\ln n +  m\ln \ln n}{n}$ based on (\ref{eq:scalinglaw-resi}); each vertical line in Figure \ref{figure5} presents the
\emph{critical} (i.e., maximal) parameter $P^*$ satisfying ${f}\cdot g \cdot  \sum_{u=d}^{K}
\frac{\binom{K}{u}\binom{P^*-K}{K-u}}{\binom{P^*}{K}} \geq
 \frac{\ln n +  m\ln \ln n}{n}$ based on (\ref{eq:scalinglaw-resi}); and each vertical line in Figure \ref{figure6} presents the
\emph{critical} parameter $f^*$ satisfying ${f^*}\cdot g \cdot  \sum_{u=d}^{K}
\frac{\binom{K}{u}\binom{P-K}{K-u}}{\binom{P}{K}} =
 \frac{\ln n +  m\ln \ln n}{n}$ based on (\ref{eq:scalinglaw-resi}). In Figures \ref{figure4}--\ref{figure6}, we also see  that the transition point of each curve is around the critical parameter illustrated by the vertical line.  Hence, Figures \ref{figure4}--\ref{figure6} are consistent with our Theorem \ref{thm:OneLaw+NodeIsolation-resi}.

To summarize, the experiments have confirmed our theoretical results in Theorem \ref{thm:OneLaw+NodeIsolation-resi}.

\section{Related Work} \label{related}

\textbf{Closely related studies.} The author's prior studies \cite{ZhaoYaganGligor,JZWiOpt15} are closely related to this paper.  Both researches  \cite{ZhaoYaganGligor,JZWiOpt15} consider the intersection of a uniform $1$-intersection graph $G_1(n, K_n,P_n)$ and an
Erd\H{o}s-R\'enyi graph $G(n,{p_n\iffalse_{on}\fi})$; specifically, \cite{ZhaoYaganGligor} presents a zero--one law for $k$-connectivity, while \cite{JZWiOpt15} improves the result of \cite{ZhaoYaganGligor} to provide the asymptotically exact probability of $k$-connectivity. To state the results of \cite{ZhaoYaganGligor,JZWiOpt15}, we let $t (K_n, P_n,1, {p_n})$ be the edge probability of $G_1(n, K_n,P_n)\bcap G(n, {p_n})$ (we use the notation $t (K_n, P_n,1, {p_n})$ since $t (K_n, P_n,d, {p_n})$ will be introduced soon and $t (K_n, P_n,1, {p_n})$ is the special case of $t (K_n, P_n,d, {p_n})$ under $d=1$).

 With the above notation, for a graph $G_1(n, K_n,P_n)\bcap G(n, {p_n})$, if there exists a sequence $\alpha_n$ such that $t (K_n, P_n,1, {p_n})   = \frac{\ln  n + (k-1) \ln \ln n   +
 {\alpha_n}}{n}$, \cite[Theorem 1]{ZhaoYaganGligor} shows that
\begin{itemize}
\item[\ding{172}] under $ \frac{{K_n}^2}{P_n} =
o(1)$ and either one of $t (K_n, P_n,1, {p_n}) = O\big(\frac{1}{n}\big)$ and $t (K_n, P_n,1, {p_n}) = o\big(\frac{1}{n}\big)$, the probability of $G_d(n, K_n,P_n)\bcap G(n, {p_n})$ being $k$-connected converges to $0$ if $\lim_{n \to \infty} \alpha_n = -\infty$, and
\item[\ding{173}] under $ \frac{K_n}{P_n} =
o(1)$ and $P_n=\Omega(n)$, the probability of $G_d(n, K_n,P_n)\bcap G(n, {p_n})$ being $k$-connected converges to $1$ if $\lim_{n \to \infty} \alpha_n = \infty$,
\end{itemize}
while  \cite[Theorem 1]{JZWiOpt15} shows that
\begin{itemize}
\item[\ding{174}]  under $ \frac{K_n}{P_n} =
o(1)$ and $P_n=\Omega(n)$, the probability of $G_d(n, K_n,P_n)\bcap G(n, {p_n})$ being $k$-connected converges to $e^{- \frac{e^{-\alpha ^*}}{(k-1)!}}$ if $\lim_{n \to \infty} \alpha_n = \alpha ^* \in (-\infty, \infty)$.
\end{itemize}

  To facilitate the comparison with \cite{JZWiOpt15,ZhaoYaganGligor}, we can also write our Theorem \ref{thm:OneLaw+NodeIsolation-resi} as $k$-connectivity result for the intersection of a uniform $d$-intersection graph $G_1(n, K_n,P_n)$ and an
Erd\H{o}s-R\'enyi graph $G(n,{p_n\iffalse_{on}\fi})$. Specifically,  (a) using the result shown later in Section \ref{sec:SystemModel} that the left hand side of (\ref{eq:scalinglaw-resi}) equals $t (K_n, P_n,d, {p_n})$, which denotes the edge probability of $G_d(n, K_n,P_n)\bcap G(n, {p_n})$, (b) \mbox{noting}  from (\ref{Gintersect3}) and the definition of $(m+1)$-connectivity that $\mathbb{P}\bigg[
\hspace{-2.9pt}\begin{array}{c}
\mathbb{G}_d(n, K_n, P_n,{f_n},g_n) \mbox{ remains connected} \\
\mbox{even after an arbitrary set of $m$ {nodes fail}.}
\end{array}\hspace{-2.9pt}
\bigg] $ is the same as $\bP{G_d(n, K_n,P_n)\bcap G(n, {p_n}) \mbox{ is $(m+1)$-connected.}}$ for $p_n := f_n \cdot g_n$, and (c) replacing $m$ in Theorem \ref{thm:OneLaw+NodeIsolation-resi} by $(k-1)$, we obtain the following result of $k$-connectivity. For a graph $G_d(n, K_n,P_n)\bcap G(n, {p_n})$, if there exists a sequence $\alpha_n$ such that $t (K_n, P_n,d, {p_n})   = \frac{\ln  n + (k-1) \ln \ln n   +
 {\alpha_n}}{n}$, where $t (K_n, P_n,d, {p_n})$ is the edge probability of $G_d(n, K_n,P_n)\bcap G(n, {p_n})$, then it holds under $ K_n =
\Omega(n^{\epsilon})$ for a positive constant $\epsilon$, $ \frac{{K_n}^2}{P_n}  =
 o\left( \frac{1}{\ln n} \right)$,  and $ \frac{K_n}{P_n} = o\left( \frac{1}{n\ln n} \right)$  that the probability of $G_d(n, K_n,P_n)\bcap G(n, {p_n})$ being $k$-connected
 \begin{itemize}
\item[\ding{202}] converges to $e^{- \frac{e^{-\alpha ^*}}{(k-1)!}}$ if $\lim_{n \to \infty} \alpha_n = \alpha ^* \in (-\infty, \infty)$,
\item[\ding{203}] converges to $1$ if $\lim_{n \to \infty} \alpha_n = \infty$,
\item[\ding{204}] converges to $0$ if $\lim_{n \to \infty} \alpha_n = -\infty$.
\end{itemize}
 Below we compare this paper and \cite{JZWiOpt15,ZhaoYaganGligor} in detail.
\begin{itemize}[leftmargin=5pt]
\item First, this paper presents the exact probability result (\ding{202}  above) in addition to a zero--one law (\ding{203} and \ding{204} above), while \cite{ZhaoYaganGligor} provides  a zero--one law (\ding{192} and \ding{193} above) and \cite{JZWiOpt15} provides  the exact probability result (\ding{194} above).
\item Second, this paper uses uniform $d$-intersection graph ${G}_d(n, K_n, P_n)$, while \cite{JZWiOpt15,ZhaoYaganGligor} uses uniform $1$-intersection graph ${G}_1(n, K_n, P_n)$ (i.e., ${G}_d(n, K_n, P_n)$ in the special case of $d=1$). More specifically, a link between two nodes in  ${G}_d(n, K_n, P_n)$ of this paper requires the sharing of at least $d$ objects, while a link between two nodes in  ${G}_1(n, K_n, P_n)$ of \cite{JZWiOpt15,ZhaoYaganGligor} requires just the sharing of at least one object (an object is a cryptographic key in \cite{JZWiOpt15,ZhaoYaganGligor}).
\item Third, this paper uses very different proof techniques compared with \cite{JZWiOpt15,ZhaoYaganGligor}. As will become clear in Section \ref{prf_idea_thm:OneLaw+NodeIsolation} later, we decompose the results into lower and upper bounds, where
\begin{itemize}
\item[(i)] the lower bound is proved by showing that our studied interest-based social network contains an Erd\H{o}s--R\'enyi graph as its spanning subgraph with probability $1-o(1)$;
\item[(ii)] the upper bound  is obtained by associating the studied $k$-connectivity property with minimum node degree.
\end{itemize}
 The above proof idea (i) associates  our studied interest-based social network with an Erd\H{o}s--R\'enyi graph, and thus can be used to establish future results for properties beyond $k$-connectivity in an interest-based social network. These results can be useful in practice. For example, combining the above association and the literature's results on the existence of a giant component in an Erd\H{o}s--R\'enyi graph, we can derive results on the existence of a giant component in an interest-based social network, where the existence of a giant component is an important property of social networks \cite{kumar2010structure} (when network-wide connectivity can be difficult, it may be desired to have that a large fraction of users belong to a connected component). Without using the above proof idea (i), the proofs in \cite{JZWiOpt15,ZhaoYaganGligor} are difficult to understand and cannot be used to obtain results for properties other than $k$-connectivity. Similar to the above proof idea (ii), the proofs in \cite{JZWiOpt15,ZhaoYaganGligor} also  associate $k$-connectivity property with minimum node degree, but the techniques and results on minimum node degree in this paper and \cite{JZWiOpt15,ZhaoYaganGligor} are different.
\end{itemize}

In addition to the above work \cite{JZWiOpt15,ZhaoYaganGligor}, graph $G_1(n, K_n,P_n)\bcap G(n, {p_n})$ has also been studied by Ya\u{g}an and Makowski \cite{yagan_onoff,makowski2015eschenauer}. For graph $G_d(n, K_n,P_n)\bcap G(n, {p_n})$, Zhao \emph{et al.} \cite{ToN17-topological,zhao2014topological} have recently studied its node degree distribution, but not connectivity.

 \iffalse

 Specifically, as will become clear in Section \ref{sec-prf-upper-bounds-details}, this paper gives the asymptotically exact probability for the property of minimum node degree being at least $k$ in $G_d(n, K_n,P_n)\bcap G(n, {p_n})$, by analyzing the asymptotically exact distribution for the number of nodes with a fixed degree. In contrast, \cite{ZhaoYaganGligor} gives only a zero--one law for the property of minimum node degree being at least $k$ in $G_1(n, K_n,P_n)\bcap G(n, {p_n})$, by bounding the first and second moments for the number of nodes with a fixed degree (the $r$th moment of a random variable $X$ is the expected value of $X^r$).
%\item Fourth, condition
\fi

\textbf{Prior work on interest-based social networks.}
 Interest-based social networks have been studied in the literature \cite{bosagh2013precision,aiello2012link,fang2014topic}, but  existing studies often  lack formal analyses (in particular for connectivity under node and link  failures). In this paper, we model an interest-based social network  by superimposing the common-interest relations over a social network, and then formally analyze its connectivity when links and nodes are allowed to fail.  Bosagh-Zadeh \textit{et al.} \cite{bosagh2013precision} introduce an interest-based framework for social networks and present conditions on the structure of the interests to achieve good
precision and recall. Aiello \cite{aiello2012link} \textit{et al.} quantify information spreading over an
interest-based online social network. Fang \cite{fang2014topic} \textit{et al.} take Flickr as the studied platform and address the problem of interest-sensitive mining.

%  \textbf{Prior work on graph intersection.} the author's prior work \cite{ZhaoYaganGligor}

\textbf{Prior work on uniform $d$-intersection graphs.}
Graph ${G}_d(n, K_n, P_n)$ models
the topology of an interest-based social network under full visibility, where full visibility
means that any pair of nodes have an edge  in between so the only requirement for a link
is the sharing of at least $d$ interests.
Graph $G_d(n, K_n, P_n)$ has been studied in the literature in terms of connectivity \cite{Bloznelis201494,GlobalSIP15-parameter} and $k$-connectivity \cite{Perfectmatchings,ANALCO}, where $k$-connectivity means that the graph remains connected despite the failure of any $(k-1)$ nodes. Other properties of $G_d(n, K_n, P_n)$ are   considered as well in the literature. For example, Bloznelis \emph{et al.}
\cite{Rybarczyk} demonstrate that a connected component with at
least a constant fraction of $n$ emerges with high probability when
the edge probability $s(K_n, P_n, d)$ exceeds $1/n$.  When $d=1$, graph ${G}_1(n, K_n, P_n)$ models
the topology of a common-interest network where two users only need to share one interest to form an edge. For ${G}_1(n, K_n, P_n)$, its ($k$)-connectivity
  has been investigated extensively \cite{ryb3,r1,yagan,DiPietroTissec,ZhaoTAC,zz}.

\textbf{Prior work on Erd\H{o}s-R\'enyi graphs.}
  Erd\H{o}s and R\'{e}nyi \cite{citeulike:4012374} %and Gilbert
%\cite{Gilbert}
 introduce the random graph model $G(n,p_n)$ defined on
a node set with size $n$ such that an edge between any two nodes
exists with probability $p_n$ {independently} of all other
edges. In a few
prior studies  \cite{citeulike:691419,etesami2016complexity,korolov2015actions,brummitt2015jigsaw}, graph $G(n,p_n)$ is used to model the topology of an online social network. For graph $G(n,p_n)$, Erd\H{o}s and R\'{e}nyi
 derive a zero-one law for connectivity in \cite{citeulike:4012374} and extend the result
 to $k$-connectivity in \cite{erdos61conn}.

 \section{Preliminaries}
\label{sec:SystemModel}

We introduce some preliminaries in this section, which will be useful for presenting the ideas to establish  Theorem \ref{thm:OneLaw+NodeIsolation-resi}  in the next section.

Clearly, the intersection of two Erd\H{o}s--R\'{e}nyi graphs defined on the same node set and generated independently is still an Erd\H{o}s--R\'{e}nyi graph. Specifically, intersecting two Erd\H{o}s--R\'{e}nyi graphs $G(n,{f_n\iffalse_{on}\fi})$ and $G(n,{g_n\iffalse_{on}\fi})$ defined on the same node set results in $G(n, {p_n})$ for
\begin{align}
p_n := f_n \cdot g_n. \label{eqnpnfngn}
\end{align}
Recall from (\ref{Gintersect3}) that $\mathbb{G}_d(n, K_n, P_n,
{f_n},g_n)   $ is the intersection of $ G_d(n, K_n,P_n) $, $G(n,{f_n\iffalse_{on}\fi})$  and $ G(n,{g_n\iffalse_{on}\fi})$.
Given the above, we know that $\mathbb{G}_d(n, K_n, P_n,
{f_n},g_n)  $ can be viewed as the intersection of $ G_d(n, K_n,P_n) $ and $ G(n,p_n)$. Then for convenience, we write $\mathbb{G}_d(n, K_n, P_n,
{f_n},g_n)$ as $\mathbb{G}_d(n, K_n,P_n, {p_n})$; in other words, we have
\begin{align}
\mathbb{G}_d(n, K_n,P_n, {p_n}):=\mathbb{G}_d(n, K_n, P_n,{f_n},g_n) \label{mathbbG_d}
\end{align}
and
\begin{align}
\mathbb{G}_d(n, K_n,P_n, {p_n}) =G_d(n, K_n,P_n)\bcap G(n, {p_n}). \label{mathbbG_dsb}
\end{align}

In $\mathbb{G}_d\iffalse_{on}\fi (n, K_n, P_n,
{p_n})$, we let the node set be $\mathcal {V}_n = \{v_1,
v_2, \ldots, v_n \}$. For each
$v_i \in \mathcal {V}_n$, the set of its $K_n$ different objects is
denoted by $S_i$, which is uniformly distributed among all
$K_n$-size subsets of a pool of $P_n$ objects.

%\linespread{1}

A uniform random $d$-intersection graph  \cite{bloznelis2013,Rybarczyk,bradonjic2010component}
 $G_d(n, K_n,P_n)$ is defined on the node set
$\mathcal {V}_n$ such that any two distinct nodes $v_i$ and $v_j$
sharing at least $d$ object(s) (an event denoted by $\Gamma_{ij}$) have
an edge in between, after each node $v_i$ independently selects $K_n$ objects \emph{uniformly at random} from the same pool $\mathcal{P}_n$ of $P_n$ objects, to form its object ring $S_i$. Clearly, event $\Gamma_{ij}$ is given by  $\big[ |S_i
\bcap S_j | \geq d \big]$, with $|A|$ denoting the
cardinality of a set $A$. In an Erd\H{o}s--R\'enyi
graph $G(n, {p_n})$ on $\mathcal {V}_n$, let  ${\Delta}_{ij}$ be the event that an edge exists
between distinct nodes $v_i$ and $v_j$. We have
$\bP{{\Delta}_{ij}} = {p_n}$, where $\mathbb{P}[\cdot]$ denotes the
probability. In graph $\mathbb{G}_d\iffalse_{on}\fi  (n, K_n, P_n,
{p_n})$, there exists an edge between nodes $v_i$ and
$v_j$ if and only if events $\Gamma_{ij}$ and ${\Delta}_{ij}$ happen at the
same time. We set event $E_{ij} : = \Gamma_{ij} \cap {\Delta}_{ij}$.
%\begin{equation}
%E_{ij} = \Gamma_{ij} \cap {\Delta}_{ij},\text{ for }1\leq i < j \leq n,
%\label{eq:E_is_S_cap_B_oy}
%\end{equation}

% Figure \ref{figure:int} presents an illustration of  uniform random $d$-intersection graph $G_d(n, K, P)$, $\text{Erd\H{o}s--R\'{e}nyi}$ graph $G(n, {p})$, and their intersection $\mathbb{G}_d\iffalse_{on}\fi  (n, K, P,
% {p})$.
%  \begin{figure*}[!t]
% \centering % \psfrag{P}{$a$}
%  \includegraphics[width=0.71\textwidth]{analco-2016-5.eps} \vspace{3pt}\caption{Uniform random $d$-intersection graphs $G_d(n, K, P)$, $\text{Erd\H{o}s--R\'{e}nyi}$ graphs $G(n, {p})$, and their intersection $\mathbb{G}_d\iffalse_{on}\fi  (n, K, P,
% {p})$.\vspace{15pt}
% }
%  \label{figure:int}
% \end{figure*}

%Note that the parameters of $\mathbb{G}_d\iffalse_{on}\fi $ include
%$n, K_n, P_n, {p_n}$ and $d$.

%Throughout the
%paper, $d$ is an arbitrary positive integer and does not scale with
%$n$.
We define
$s(K_n, P_n,d) $ and $t (K_n, P_n,d, {p_n})$ as the edge probabilities of graphs
$G_d(n, K_n, P_n)$ and $\mathbb{G}_d(n,
K_n, P_n, {p_n})$, respectively.  From $E_{ij} = \Gamma_{ij} \cap {\Delta}_{ij}$ and the
independence of $ \Gamma_{i j} $ and ${\Delta}_{ij} $, we obtain
\begin{align}
& {t (K_n, P_n,d, {p_n})} \nonumber \\ &  =  \mathbb{P} [E_{i j} ]  =  \mathbb{P} [{\Delta}_{ij} ]
\cdot \mathbb{P} [\Gamma_{i j} ] =  {p_n}\iffalse_{on}\fi \cdot
s(K_n, P_n,d). \label{eq_pre}
\end{align}

 By definition, $s(K_n, P_n,d)$ is determined through
\begin{align}
s(K_n, P_n,d) & =  \mathbb{P} [\Gamma_{i j} ] = \sum_{u=d}^{K_n}
  \mathbb{P}[|S_{i} \cap S_{j}| = u] , \label{psq1}
\end{align}
where we derive $\mathbb{P}[|S_{i} \cap S_{j}| = u]$ as follows.

We observe that $S_{i}$ and $S_{j}$ are independently and uniformly
selected from all $K_n$-size subsets of an object pool of size $P_n$.
Under $(|S_{i} \cap S_{j}| = u)$, after $S_i$ is determined, $S_j$
is constructed by selecting $u$ objects out of $S_i$ and $(K_n-u)$ objects
out of the object pool $\mathcal {P}_n$. Hence, if $P_n \geq 2K_n$, we have
\begin{align}
& \mathbb{P}[|S_{i} \cap S_{j}| = u]  =
\frac{\binom{K_n}{u}\binom{P_n-K_n}{K_n-u}}{\binom{P_n}{K_n}},
 \quad\quad\text{for } u =0,1,\ldots, K_n,
\label{u3}
\end{align}
which along with (\ref{eq_pre}) and (\ref{psq1}) yields
\begin{align}
t (K_n, P_n,d, {p_n}) = {p_n}\iffalse_{on}\fi \cdot  \sum_{u=d}^{K_n}
\frac{\binom{K_n}{u}\binom{P_n-K_n}{K_n-u}}{\binom{P_n}{K_n}}. \label{u4}
\end{align}
Substituting (\ref{eqnpnfngn}) into (\ref{u4}), we have
\begin{align}
 {t (K_n, P_n,d, {p_n})} &=  {f_n}\iffalse_{on}\fi \cdot g_n \cdot  \sum_{u=d}^{K_n}
\frac{\binom{K_n}{u}\binom{P_n-K_n}{K_n-u}}{\binom{P_n}{K_n}}. \label{eq_preeqnpnfngn}
\end{align}

  \section{Basic Ideas to Establish Theorem \ref{thm:OneLaw+NodeIsolation-resi}}  \label{prf_idea_thm:OneLaw+NodeIsolation}

  In this section, we present the basic ideas to establish  Theorem \ref{thm:OneLaw+NodeIsolation-resi}.

  \subsection{An Equivalent Form of Theorem \ref{thm:OneLaw+NodeIsolation-resi}}

From the definition of $(m+1)$-connectivity \cite{Citeulike:505396}, connectivity is still preserved after an arbitrary set of $m$ nodes fail, if and only if the original graph is  $(m+1)$-connected. Given this and the fact that the left hand side of (\ref{eq:scalinglaw-resi}) equals $t (K_n, P_n,d, {p_n})$ from (\ref{eq_preeqnpnfngn}), Theorem \ref{thm:exact_qcomposite-kcon} below is equivalent with Theorem \ref{thm:OneLaw+NodeIsolation-resi}. Note that we first replace $m$ in Theorem \ref{thm:OneLaw+NodeIsolation-resi} by $(k-1)$ and then rephrase the result as Theorem \ref{thm:exact_qcomposite-kcon}.

\begin{thm}\label{thm:exact_qcomposite-kcon}

For a graph $\mathbb{G}_d(n, K_n,P_n, {p_n})$, with a sequence $\alpha_n$ defined by
\begin{align}
t (K_n, P_n,d, {p_n})  & = \frac{\ln  n + (k-1) \ln \ln n   +
 {\alpha_n}}{n}  , \label{peq1sbsc-kcon}
\end{align}
then it holds under $ K_n =
\Omega(n^{\epsilon})$ for a positive constant $\epsilon$, $ \frac{{K_n}^2}{P_n}  =
 o\left( \frac{1}{\ln n} \right)$,  and $ \frac{K_n}{P_n} = o\left( \frac{1}{n\ln n} \right)$  that
\iffalse
\begin{subnumcases}
{ \hspace{-7pt}  \lim_{n \rightarrow \infty }\hspace{-2pt} \mathbb{P}\hspace{-1.5pt}\bigg[
\hspace{-4.9pt}\begin{array}{c}
\mathbb{G}_d(n, K_n,P_n, {p_n}) \\
\mbox{is $k$-connected.}
\end{array}\hspace{-4.9pt}
\bigg] \hspace{-3.7pt}=\hspace{-3.7pt}}  \hspace{-5pt}0,\quad\hspace{-9pt}\text{if  }\hspace{-5pt}\lim_{n \to \infty}\hspace{-1.5pt}{\alpha_n}  \hspace{-2.5pt} =  \hspace{-2.5pt} - \infty, \label{thm-con-eq-0} \\
\hspace{-5pt}1,\quad\hspace{-9pt}\text{if  }\hspace{-5pt}\lim_{n \to \infty}\hspace{-1.5pt}{\alpha_n}    \hspace{-2.5pt}=  \hspace{-2.5pt}  \infty, \label{thm-con-eq-1}
\end{subnumcases}
\fi
\begin{align}
& \hspace{-35pt}  \lim_{n \rightarrow \infty } \bP{\mathbb{G}_d(n, K_n,P_n, {p_n}) \mbox{ is $k$-connected.}}
%& \hspace{-57pt} \lim_{n \rightarrow \infty }  \mathbb{P} \bigg[
%\hspace{-3pt}\begin{array}{c}
%\mathbb{G}_q(n, K_n,P_n, {p_n}) \\
%\mbox{is $k$-connected.}
%\end{array}\hspace{-3pt}
%\bigg]
 \nonumber
% \\  & \hspace{-27pt}
% =   e^{- \frac{e^{-\lim_{n \to \infty}{\alpha_n}}}{(k-1)!}}  \label{thm-con-eq-compact}
\\ &  \hspace{-35pt}  =  e^{- \frac{e^{-\lim_{n \to \infty} \alpha_{_n}}}{(k-1)!}} \label{thm-mnd-alpha-finite-kcon-compact}
\end{align}
\vspace{-5pt}
\begin{subnumcases}{=}
e^{- \frac{e^{-\alpha ^*}}{(k-1)!}}, &\textrm{if } $\lim\limits_{n \to \infty} \alpha_n = \alpha ^* \in (-\infty, \infty)$ ,\label{thm-mnd-alpha-finite-kcon} \\
1, & \textrm{if } $\lim\limits_{n \to \infty} \alpha_n =  \infty$,\label{thm-mnd-alpha-infinite-kcon}\\
0, & \textrm{if } $\lim\limits_{n \to \infty} \alpha_n = - \infty$. \label{thm-mnd-alpha-minus-infinite-kcon}
\end{subnumcases}

\end{thm}

In the rest of the paper, we will discuss how to establish Theorem \ref{thm:exact_qcomposite-kcon}. Proving Theorem \ref{thm:exact_qcomposite-kcon} is equivalent to  proving Theorem \ref{thm:OneLaw+NodeIsolation-resi}.

\textbf{Interpreting Theorem \ref{thm:exact_qcomposite-kcon}.} For $k$-connectivity in
$\mathbb{G}_d(n, K_n, P_n,{f_n},g_n)$, the result (\ref{thm-mnd-alpha-finite-kcon-compact}) of Theorem \ref{thm:exact_qcomposite-kcon} presents the asymptotically exact
probability, while (\ref{thm-mnd-alpha-infinite-kcon}) and (\ref{thm-mnd-alpha-minus-infinite-kcon}) of Theorem \ref{thm:exact_qcomposite-kcon} together constitute a zero--one law, where a zero--one law means that the probability of a graph having a certain property asymptotically converges to $0$ under some conditions and to $1$ under some other conditions.  The result (\ref{thm-mnd-alpha-finite-kcon-compact}) compactly summarize (\ref{thm-mnd-alpha-finite-kcon})--(\ref{thm-mnd-alpha-minus-infinite-kcon}).

To demonstrate Theorem \ref{thm:exact_qcomposite-kcon}, we explain the basic ideas in the rest of the section, and provide additional details in the Appendix. For simplicity, below we often write $\mathbb{G}_d(n, K_n,P_n, {p_n})$ as $\mathbb{G}_{n,d}\iffalse_{on}\fi$, ${t (K_n, P_n,d, {p_n})} $ in (\ref{eq_pre}) as $t_{n,d}$, and $s(K_n, P_n,d)$ in (\ref{psq1}) as $s_{n,d}$. Then  (\ref{eq_pre}) means
\begin{align}
t_{n,d}= s_{n,d} {p_n} . \label{eq_premean}
\end{align}

 {\color{black}
 \subsection{Basic ideas for proving Theorem \ref{thm:exact_qcomposite-kcon}} \label{section-basicprf}

}

% First, we note that

% Theorems \ref{thm:exact_qcomposite-kcon}--\ref{thm-pm} uses the condition (\ref{scalingP-stronger}) (i.e., $P_n = \Omega(n)$ for $q=1$, and $P_n = \omega\big(n^{2-\frac{1}{q}}(\ln n)^{2+\frac{1}{q}}\big)$ for $q\geq 2$).

The basic ideas to show Theorem \ref{thm:exact_qcomposite-kcon} are as follows. We decompose the theorem results into lower and upper bounds, where the lower bound  is proved by associating our studied uniform $d$-intersection graph with an Erd\H{o}s--R\'enyi graph, while the upper bound  is obtained by associating the studied $k$-connectivity property in Theorem \ref{thm:exact_qcomposite-kcon} with minimum node degree.

\subsubsection{\textbf{Decomposing the results into lower and upper bounds}} \label{sec-Decomposing}

Note that in Theorem \ref{thm:exact_qcomposite-kcon}, the results (\ref{thm-mnd-alpha-finite-kcon})--(\ref{thm-mnd-alpha-minus-infinite-kcon}) are compactly summarized as (\ref{thm-mnd-alpha-finite-kcon-compact}); i.e., $\lim\limits_{n \to \infty}\bP{\text{$\mathbb{G}_{n,d}\iffalse_{on}^{(d)}\fi$  is $k$-connected.}}=e^{- \frac{e^{-\lim_{n \to \infty} \alpha_n }}{(k-1)!}}$. To prove (\ref{thm-mnd-alpha-finite-kcon-compact}) via decomposition, we show that the probability $\mathbb{P}[\hspace{2pt}\mathbb{G}_{n,d}\text{ is
  $k$-connected.}\hspace{2pt}]$ has a lower bound $e^{-
\frac{e^{- \iffalse \lim\limits_ \fi \lim_{n \to \infty}{\alpha_n}}}{(k-1)!}} \times [1 -o(1)]$ and an upper bound $e^{-
\frac{e^{- \iffalse \lim\limits_ \fi \lim_{n \to \infty}{\alpha_n}}}{(k-1)!}} \times [1 + o(1)]$, where a sequence $x_n$ can be written as $o(1)$ if $\lim_{n \to \infty} x_n = 0$. Afterwards, the obtained (\ref{thm-mnd-alpha-finite-kcon-compact}) implies (\ref{thm-mnd-alpha-finite-kcon})--(\ref{thm-mnd-alpha-minus-infinite-kcon})

\subsubsection{\textbf{Proving the lower bound by showing that our interest-based social network $\mathbb{G}_{n,d}$ contains an Erd\H{o}s--R\'enyi graph}} \label{sec-prf-lower-bounds}

To prove the lower bound of $k$-connectivity in our studied interest-based social network $\mathbb{G}_{n,d}$, we will show that the studied network $\mathbb{G}_{n,d}$ contains an Erd\H{o}s--R\'enyi graph as its spanning subgraph with probability $1-o(1)$, and   show that the lower bound also holds for the Erd\H{o}s--R\'enyi graph. More specifically, the Erd\H{o}s--R\'enyi graph under the corresponding conditions is $k$-connected with probability \mbox{$e^{-
\frac{e^{- \iffalse \lim\limits_ \fi \lim_{n \to \infty}{\alpha_n}}}{(k-1)!}} \times [1-o(1)]$} .

We provide more details for the above idea in Section \ref{sec-prf-lower-bounds-details}.

\subsubsection{\textbf{Proving the upper bound by considering minimum node degree}} \label{sec-prf-upper-bounds}~

To prove the upper bound of $k$-connectivity in our studied interest-based social network $\mathbb{G}_{n,d}$, we leverage the necessary condition on the minimum (node) degree enforced by $k$-connectivity, and explain that the upper bound also holds for the requirement of the minimum degree. Specifically, because a necessary condition for a graph to be $k$-connected is that the minimum degree is at least $k$ \cite{Citeulike:505396}, $\bP{\text{$\mathbb{G}_{n,d}\iffalse_{on}^{(d)}\fi$  has a minimum degree at least $k$.}}$ provides an upper bound for $\bP{\text{$\mathbb{G}_{n,d}\iffalse_{on}^{(d)}\fi$  is $k$-connected.}}$. We will prove that $\bP{\text{$\mathbb{G}_{n,d}\iffalse_{on}^{(d)}\fi$  has a minimum degree at least $k$.}}$ is upper bounded by $e^{-
\frac{e^{- \iffalse \lim\limits_ \fi \lim_{n \to \infty}{\alpha_n}}}{(k-1)!}} \times [1 + o(1)]$ so it becomes immediately clear  that $\bP{\text{$\mathbb{G}_{n,d}\iffalse_{on}^{(d)}\fi$  is $k$-connected.}}$ is also upper bounded by $e^{-
\frac{e^{- \iffalse \lim\limits_ \fi \lim_{n \to \infty}{\alpha_n}}}{(k-1)!}} \times [1 + o(1)]$.

We give more details for the above idea in Section \ref{sec-prf-upper-bounds-details}.

In addition to the arguments above, we also find it useful to confine the deviation  $\alpha_n$ in Theorem  \ref{thm:exact_qcomposite-kcon}. We discuss this idea as follows.

\subsubsection{\mbox{Confining the deviation  $\alpha_n$ in Theorem  \ref{thm:exact_qcomposite-kcon}}} \label{sec-Confining}~

We will show that to prove Theorem \ref{thm:exact_qcomposite-kcon}, the deviation $\alpha_n$ in the theorem statement can  be confined as $\pm  o(\ln n)$. More specifically, if Theorem \ref{thm:exact_qcomposite-kcon} holds under the extra condition $|\alpha_n |= o(\ln n)$, then Theorem \ref{thm:exact_qcomposite-kcon} also holds regardless of the extra condition. This extra condition will be useful for the aforementioned steps in Sections \ref{sec-prf-lower-bounds} and \ref{sec-prf-upper-bounds}. We present more details for the above idea in Appendix D of the full version \cite{fullpdfaaaia}.

\iffalse

We prove Lemma \ref{lem-mnd} in Appendix ?? of the full version \cite{fullpdfaaaia} due to space limitation.

We present more details for the above idea in Section \ref{sec-Confining-details}. \fi

% Confining the deviation from the critical scaling

% We can introduce the following extra conditions to establish the theorems:
% \begin{itemize}
% \item $|\alpha_n |= o(\ln n)$ for proving Theorem \ref{thm:exact_qcomposite-kcon},
% \item $|\beta_n |= o(\ln n)$ for proving Theorem \ref{thm-krob},
% \item  $|\gamma_n |= o(\ln n)$ for proving Theorem \ref{thm-hc}, and
% \item  $|\xi_n |= o(\ln n)$ for proving Theorem \ref{thm-pm}.
% \end{itemize}

% extra condition $|\xi_n |= o(\ln n)$

\subsection{More details for proving the lower bound of   Section \ref{sec-Decomposing}} \label{sec-prf-lower-bounds-details}

The idea to prove the lower bound $e^{-
\frac{e^{- \iffalse \lim\limits_ \fi \lim_{n \to \infty}{\alpha_n}}}{(k-1)!}} \times [1 - o(1)]$ for $\mathbb{P}[\hspace{2pt}\mathbb{G}_{n,d}\text{ is
  $k$-connected.}\hspace{2pt}]$ has been explained in Section \ref{sec-prf-lower-bounds}. As explained, we associate the studied interest-based social network $\mathbb{G}_{n,d}$ with an Erd\H{o}s--R\'enyi graph $G(n,z_n)$. The result is given as Lemma \ref{lem-cp_rig_er} below.

Before presenting Lemma \ref{lem-cp_rig_er}, we first discuss the notion of graph coupling which
associates two random graphs together. The goal is to convert a problem in one random graph to the corresponding problem in another random graph, in order to solve the original problem.
Formally, a coupling
\cite{zz,2013arXiv1301.0466R,Krzywdzi} of two random graphs
$G_1$ and $G_2$ means a probability space on which random graphs
$G_1'$ and $G_2'$ are defined such that $G_1'$ and $G_2'$ have the
same distributions as $G_1$ and $G_2$, respectively. For notation
brevity, we simply say $G_1$ is a spanning subgraph\footnote{A graph
$G_a$ is a spanning subgraph (resp., spanning supergraph) of a graph
$G_b$ if $G_a$ and $G_b$ have the same node set, and the edge set of
$G_a$ is a subset (resp., superset) of the edge set of $G_b$. \label{ft-span}}
(resp., spanning supergraph) of $G_2$ if $G_1'$ is a spanning
subgraph of $G_2'$.

 Following Rybarczyk's notation \cite{zz}, we
write
\begin{align}
G_1 \succeq & G_2 \quad (\textrm{resp.}, G_1 \succeq_{1-o(1)} G_2)
\label{g1g2coupling}
\end{align}
if there exists a coupling under which $G_2$ is a spanning subgraph
of $G_1$ with probability $1$ (resp., $1-o(1)$); i.e., $G_1$ is a spanning supergraph
of $G_2$ with probability $1$ (resp., $1-o(1)$), where the notions of spanning subgraph and supergraph have been defined in Footnote \ref{ft-span}.

Given the above notation, we now present Lemma \ref{lem-cp_rig_er}, which couples our graph   $\mathbb{G}_{n,d}\iffalse_{on}^{(d)}\fi$ with an Erd\H{o}s--R\'enyi graph.

 \iffalse

 Lemma \ref{lem-only-prove-lnlnn} below ?? uses the notion of graph coupling \cite{zz,2013arXiv1301.0466R}.
A coupling of two random graphs $G_1$ and
$G_2$ means a probability space on which random graphs $G_1'$ and
$G_2'$ are defined such that $G_1'$ and $G_2'$ have the same
distributions as $G_1$ and $G_2$, respectively. If $G_1'$ is a spanning subgraph
(resp., spanning supergraph) of $G_2'$, we say that under the coupling, $G_1$ is a spanning subgraph
(resp.,  spanning supergraph) of $G_2$, which yields that for any monotone increasing property $\mathcal {I}$, the probability of $G_1$ having $\mathcal {I}$ is at most (resp., at least) the probability of $G_2$ having $\mathcal {I}$.

\fi

\begin{lem} \label{lem-cp_rig_er}
 If $ K_n =
\Omega(n^{\epsilon})$ for a positive constant $\epsilon$, \vspace{1pt} $ \frac{{K_n}^2}{P_n}  =
 o\left( \frac{1}{\ln n} \right)$, $ \frac{K_n}{P_n} = o\left( \frac{1}{n\ln n} \right)$, and $\frac{{K_n}^2}{P_n} = \omega\big(\frac{(\ln n)^6}{n^2}\big)$, then there exists a sequence $z_n$ satisfying
\begin{align}
\textstyle{z_n = t_{n,d} \times \big[1-o\big(\frac{1}{\ln n}\big)\big]} \label{ERgraph-sn-defn}
\end{align}
 such that
 \begin{align}
\mathbb{G}_{n,d}  \succeq_{1-o(1)}G(n,z_n);\label{lem1-rescp}
\end{align}
i.e., graph $\mathbb{G}_{n,d}\iffalse_{on}^{(d)}\fi$ contains an Erd\H{o}s--R\'{e}nyi graph $G(n,z_n)$ as a spanning subgraph with probability $1-o(1)$ (when we couple the two graphs on the same probability space and define them on the same node set), where we note that $t_{n,d}$ is the edge probability of $\mathbb{G}_{n,d}\iffalse_{on}^{(d)}\fi$, and $z_n$ is the edge probability of $G(n,z_n)$.
 \end{lem}
%  Due to space limitation, we prove Lemma \ref{lem-cp_rig_er} in
%  Appendix \ref{appsec-lem-cp_rig_er} of the full version \cite{fullpdfaaaia}.

\begin{rem}
From \cite{zz}, since $k$-connectivity is a monotone increasing graph property, (\ref{ERgraph-sn-defn}) further implies
% (a graph property is called
% monotone increasing if it holds under the addition of edges in a
% graph),
\begin{align}
 \bP{\text{$\mathbb{G}_{n,d}\iffalse_{on}^{(d)}\fi$  is $k$-connected.}}   \geq \bP{\text{$G(n,z_n)$ is $k$-connected.}} - o(1) . \label{Gq-kcon-lower-bound-tonsb}
\end{align}
 \end{rem}

Recall from (\ref{mathbbG_dsb}) that $\mathbb{G}_{n,d}$ (i.e., $\mathbb{G}_d(n, K_n,P_n, {p_n})$) is the intersection of a uniform $d$-intersection graph $G_d(n, K_n,P_n)$ and an
Erd\H{o}s-R\'enyi graph $G(n,{p_n\iffalse_{on}\fi})$. To prove Lemma \ref{lem-cp_rig_er} which associates $\mathbb{G}_{n,d}$ with an
Erd\H{o}s-R\'enyi graph, we   establish Lemma \ref{lem-cpgraph-rigrig} below which couples $G_d(n, K_n,P_n)$ with another
Erd\H{o}s-R\'enyi graph.

\begin{lem} \label{lem-cpgraph-rigrig}
 If $ K_n =
\Omega(n^{\epsilon})$ for a positive constant $\epsilon$, \vspace{1pt} $ \frac{{K_n}^2}{P_n}  =
 o\left( \frac{1}{\ln n} \right)$, $ \frac{K_n}{P_n} = o\left( \frac{1}{n\ln n} \right)$, and $\frac{{K_n}^2}{P_n} = \omega\big(\frac{(\ln n)^6}{n^2}\big)$, then there exists a sequence $y_n$ satisfying
\begin{align}
\textstyle{y_n = s_{n,d} \times \big[1-o\big(\frac{1}{\ln n}\big)\big]} \label{ERgraph-sn-defn-reduced}
\end{align}
such that
\begin{align}
G_d(n,K_n,P_n)   \succeq_{1-o(1)}G(n,y_n); \label{lem1-rescp}
\end{align}
i.e., a uniform $d$-intersection graph $G_d(n, K_n,P_n)$ contains an Erd\H{o}s--R\'{e}nyi graph $G(n,y_n)$ as a spanning subgraph with probability $1-o(1)$ (when we couple the two graphs on the same probability space and define them on the same node set), where $s_{n,d}$ is the edge probability of $G_d(n, K_n, P_n)$.
 \end{lem}

We will discuss the proof of Lemma \ref{lem-cpgraph-rigrig} later. Below we show that Lemma \ref{lem-cp_rig_er} follows from Lemma \ref{lem-cpgraph-rigrig}.

\textbf{Proof of Lemma \ref{lem-cp_rig_er} using Lemma \ref{lem-cpgraph-rigrig}:}

As noted in Lemmas  \ref{lem-cp_rig_er} and \ref{lem-cpgraph-rigrig}, we will couple different random graphs together. We now use Lemma \ref{lem-cpgraph-rigrig} to prove Lemma \ref{lem-cp_rig_er}.

%
%We recall from (\ref{mathbbG_d}) that
%\begin{equation}
%\mathbb{G}_{n,d}\iffalse_{on}\fi  = G_d(n, K_n, P_n) \cap G(n, p_n).
% \label{eq:G_on_is_RKG_cap_ER_oyton-reduced}
%\end{equation}
%After intersecting $G_d(n, K_n, P_n)$ (resp., $G(n,y_n)$) with $G(n, p_n)$, we obtain $G_d(n, K_n, P_n) \cap G(n, p_n)$ (resp., $G(n,y_n) \cap G(n, p_n)$), where $G_d(n, K_n, P_n) \cap G(n, p_n)$ is $\mathbb{G}_{n,d}$ from (\ref{eq:G_on_is_RKG_cap_ER_oyton-reduced}), and $G(n,y_n) \cap G(n, p_n)$ becomes an Erd\H{o}s--R\'{e}nyi graph $G(n,y_n p_n)$.
%

From (\ref{lem1-rescp}) of Lemma \ref{lem-cpgraph-rigrig}, $G_d(n, K_n, P_n)$ contains an Erd\H{o}s--R\'{e}nyi graph $G(n,y_n)$ as a spanning subgraph with probability $1-o(1)$ for $y_n$ in (\ref{ERgraph-sn-defn-reduced}), when we couple the two graphs on the same probability space and define them on the same node set (say $\mathcal{V}$). Let $\mathcal{E}_d$ be the edge set of $G_d(n, K_n, P_n)$, and $\mathcal{E}_y$ be the edge set of $G(n,y_n)$, so that $\mathcal{E}_y\subseteq \mathcal{E}_d$ with probability $1-o(1)$. We further define an Erd\H{o}s--R\'{e}nyi graph $G(n,p_n)$ on the same node set $\mathcal{V}$ and denote its edge set by $\mathcal{E}_p$. Then we intersect $G_d(n, K_n, P_n)$ and $G(n,p_n)$ to obtain $G_d(n, K_n,P_n)\bcap G(n, {p_n})$ defined on the node set $\mathcal{V}$ and the edge set   $\mathcal{E}_d \bcap \mathcal{E}_p$. We also intersect $G(n,y_n)$ and $G(n,p_n)$ to obtain $G(n,y_n)\bcap G(n,p_n)$
defined on the node set $\mathcal{V}$ and the edge set   $\mathcal{E}_y \bcap \mathcal{E}_p$. Given the above $\mathcal{E}_y\subseteq \mathcal{E}_d$ with probability $1-o(1)$, we have $\mathcal{E}_y\bcap \mathcal{E}_p\subseteq \mathcal{E}_d\bcap \mathcal{E}_p$ with probability $1-o(1)$. This means that
% , and intersecting an Erd\H{o}s--R\'{e}nyi graph $G(n,y_n)$ with an Erd\H{o}s--R\'{e}nyi graph $G(n,p_n)$ to obtain $G(n,y_n)\bcap G(n,p_n)$, from (\ref{lem1-rescp}) and the definition of spanning subgraph, we can obtain that
 $G_d(n, K_n,P_n)\bcap G(n, {p_n})$ contains $G(n,y_n)\bcap G(n,p_n)$ as a spanning subgraph with probability $1-o(1)$. Put formally, we have
\begin{align}
  G_d(n, K_n,P_n)\bcap G(n, {p_n}) \succeq_{1-o(1)} G(n,y_n)\bcap G(n,p_n).
\label{g1g2coupling-v2c}
\end{align}
We can also view (\ref{g1g2coupling-v2c}) as the result of
intersecting each side of (\ref{lem1-rescp}) with an Erd\H{o}s--R\'{e}nyi graph $G(n,p_n)$.
We now analyze both sides of (\ref{g1g2coupling-v2c}). First,  we obtain from   (\ref{mathbbG_dsb}) that
\begin{align}
\mathbb{G}_{n,d}  =G_d(n, K_n,P_n)\bcap G(n, {p_n}),
\label{g1g2coupling-v2a}
\end{align}
where we recall that $\mathbb{G}_{n,d}$ is short for $\mathbb{G}_d(n, K_n,P_n, {p_n})$.

Second, the intersection of an Erd\H{o}s--R\'{e}nyi graph $G(n,y_n)$ and an Erd\H{o}s--R\'{e}nyi graph $G(n,p_n)$ will induce an Erd\H{o}s--R\'{e}nyi graph $G(n,y_n p_n)$. This is straightforward from the definition of an Erd\H{o}s--R\'{e}nyi graph $G(n,a_n)$: any node pair among $n$ nodes have an edge in between independently with the same probability $a_n$. Hence, we have
\begin{align}
G(n,y_n p_n)  =G(n,y_n)\bcap G(n,p_n).
\label{g1g2coupling-v2b}
\end{align}

Substituting (\ref{g1g2coupling-v2a}) and (\ref{g1g2coupling-v2b}) into (\ref{g1g2coupling-v2c}), we obtain \begin{align}
\mathbb{G}_{n,d} \succeq_{1-o(1)} G(n,y_n p_n);
\label{g1g2coupling-v2}
\end{align}
 i.e., $\mathbb{G}_{n,d}$ contains an Erd\H{o}s--R\'{e}nyi graph $G(n,y_n p_n)$ as a spanning subgraph with probability $1-o(1)$ for $y_n$ in (\ref{ERgraph-sn-defn-reduced}) (when we couple the two graph intersections on the same probability space and define them on the same node set).
We further analyze $y_n p_n$ in (\ref{g1g2coupling-v2}). From (\ref{eq_premean}) (i.e., $t_{n,d} = s_{n,d} p_n$) and (\ref{ERgraph-sn-defn-reduced}), it follows that
\begin{align}
y_n p_n & = s_{n,d} \times \textstyle{\big[1-o\big(\frac{1}{\ln n}\big)\big]}  \times  p_n =   t_{n,d} \times \textstyle{\big[1-o\big(\frac{1}{\ln n}\big)\big]}. \label{g1g2coupling-v2znnew}
\end{align}
Given (\ref{g1g2coupling-v2}) and (\ref{g1g2coupling-v2znnew}), we set $z_n$ in (\ref{ERgraph-sn-defn}) as $y_n p_n$, and thus complete proving  Lemma \ref{lem-cp_rig_er} using Lemma \ref{lem-cpgraph-rigrig}. \qeda

%Once we show $z_n$ in (\ref{ERgraph-sn-defn}) can be set as $y_n p_n$, the proof of Lemma \ref{lem-cp_rig_er} will be completed ???.

\textbf{Basic Ideas of Proving Lemma \ref{lem-cpgraph-rigrig}:}

To establish Lemma \ref{lem-cpgraph-rigrig}, we explain the basic ideas here and present the formal proof   in Appendix \ref{app-prf-lem-cpgraph-rigrig}.
The proof of Lemma \ref{lem-cpgraph-rigrig} is quite involved, since uniform $d$-intersection graph $G_d(n, K_n,P_n)$ and Erd\H{o}s--R\'{e}nyi graph $G(n,y_n)$ associated by Lemma \ref{lem-cpgraph-rigrig} are very different. For instance, while edges in $G(n,y_n)$ are all independent, not all edges in $G_d(n, K_n,P_n)$ are  independent with each other, since the event that nodes $v_1$ and $v_2$ share at least $d$ objects to have an edge in between, and the event that nodes $v_1$ and $v_3$ share at least $d$ objects to have an edge in between, may induce higher chance for  the event that nodes $v_2$ and $v_3$ share at least $d$ objects to have an edge in between.

 To prove Lemma \ref{lem-cpgraph-rigrig}, we
introduce an auxiliary graph called the \emph{binomial
$d$-intersection
 graph} $H_d(n,x_n,P_n)$ \cite{Rybarczyk,Assortativity,bloznelis2013}, which
 can be defined on $n$ nodes by the following process.
  There exists an object pool of size $P_n$. Each object in the
pool is added to each node {independently} with probability
$x_n$. After each node obtains a set of objects, two nodes establish an edge in between if and only if they share at least $d$ objects. Clearly, the only difference between binomial $d$-intersection
 graph $H_d(n,x_n,P_n)$ and uniform $d$-intersection   graph $G_d(n,K_n,P_n)$ is that in the former,
  the number of objects assigned to each
 node obeys a binomial distribution with $P_n$ as
the number of trials, and with $x_n$ as the success probability in
each trial, while in the latter graph, such number equals $K_n$ with
probability $1$.

To prove Lemma \ref{lem-cpgraph-rigrig}, we present Lemmas \ref{brig_urig} and \ref{er_brig} below. Lemma \ref{brig_urig} shows
$G_d(n,K_n,P_n)  \succeq_{1-o(1)}H_d(n,x_n,P_n)$; i.e.,    a uniform $d$-intersection graph $G_d(n, K_n,P_n)$ contains a binomial $d$-intersection
 graph $H_d(n,x_n,P_n)$ as a spanning subgraph with probability $1-o(1)$ (when we couple the two graphs on the same probability space and define them on the same node set). Lemma \ref{er_brig} shows $H_d(n,x_n,P_n) \succeq_{1-o(1)} G(n,y_n)$; i.e.,    a binomial $d$-intersection
 graph $H_d(n,x_n,P_n)$ contains an Erd\H{o}s--R\'{e}nyi graph $G(n,y_n)$ as a spanning subgraph with probability $1-o(1)$ (when we couple the two graphs on the same probability space and define them on the same node set). Then via a transitive argument \cite[Fact
3]{2013arXiv1301.0466R}, we obtain $G_d(n,K_n,P_n)   \succeq_{1-o(1)}G(n,y_n)$; i.e.,    a uniform $d$-intersection graph $G_d(n, K_n,P_n)$ contains an Erd\H{o}s--R\'{e}nyi graph $G(n,y_n)$ as a spanning subgraph with probability $1-o(1)$ (when we couple the two graphs on the same probability space and define them on the same node set). Of course, we still need to show that (i) given the conditions of Lemma \ref{lem-cpgraph-rigrig}, all
conditions in Lemmas \ref{brig_urig} and \ref{er_brig} hold; and (ii)  $y_n$ defined in Lemma \ref{er_brig} (specifically (\ref{pnpb01})) satisfies Lemma \ref{lem-cpgraph-rigrig}  (specifically (\ref{ERgraph-sn-defn-reduced})). We provide the formal details of proving Lemma \ref{lem-cpgraph-rigrig} in Appendix \ref{app-prf-lem-cpgraph-rigrig}.

% $ \frac{{K_n}^2}{P_n}  =
%  o\left( \frac{1}{\ln n} \right)$, $ \frac{K_n}{P_n} = o\left( \frac{1}{n\ln n} \right)$, \vspace{1pt} $ K_n =
% \Omega(n^{\epsilon})$ for a positive constant $\epsilon$, and $\frac{{K_n}^2}{P_n} = \omega\big(\frac{1}{n^2}\big)$,

% \begin{lem} \label{brig_urig}
% If $ K_n =
% \Omega(n^{\epsilon})$ for a positive constant $\epsilon$, and $ \frac{{K_n}^2}{P_n}  =
% o\left( \frac{1}{\ln n} \right)$, with
% $x_n$ set by
% \begin{align}
%  x_n   = \textstyle{\frac{K_n}{P_n}
%  \Big(1 - \sqrt{\frac{3\ln
% n}{K_n }}\hspace{2pt}\Big)}, \label{pnKn}
%  \end{align}
% then %the edge probability $p_b$ in binomial $d$-intersection
% % graph $H_d(n,x_n,P_n)$ satisfies
% %\begin{align}
% %p_b & =  \textstyle{ \frac{1}{d!} \big( \frac{{K_n}^2}{P_n}
% %\big)^{d}} \cdot [1\pm o(1)], \label{pbps01}
% %\end{align}
% %and
%  it holds that
% \begin{align}
%   G_d(n,K_n,P_n) & \succeq_{1-o(1)}H_d(n,x_n,P_n). \label{eq_brig_urig}
% \end{align}

% \end{lem}

Lemmas \ref{brig_urig} and \ref{er_brig} below are useful for proving Lemma \ref{lem-cpgraph-rigrig}.

\begin{lem} \label{brig_urig}
If $K_n = \omega(\ln n)$ and $ \frac{{K_n}^2}{P_n}  =
 o\left( 1\right)$, with
$x_n$ set by
\begin{align}
 x_n   = \textstyle{\frac{K_n}{P_n}
 \Big(1 - \sqrt{\frac{3\ln
n}{K_n }}\hspace{2pt}\Big)}, \label{pnKn}
 \end{align}
then %the edge probability $p_b$ in binomial $d$-intersection
% graph $H_d(n,x_n,P_n)$ satisfies
%\begin{align}
%p_b & =  \textstyle{ \frac{1}{d!} \big( \frac{{K_n}^2}{P_n}
%\big)^{d}} \cdot [1\pm o(1)], \label{pbps01}
%\end{align}
%and
 it holds that
\begin{align}
  G_d(n,K_n,P_n) & \succeq_{1-o(1)}H_d(n,x_n,P_n). \label{eq_brig_urig}
\end{align}

\end{lem}

\begin{lem} \label{er_brig}
If
 \begin{align}
\textstyle{ x_n  P_n } & = \Omega(n^{\epsilon}) \text{ for a positive constant $\epsilon$}, \label{er_brig-eq1} \\  \textstyle{ {x_n} } & = \textstyle{o\left( \frac{1}{n\ln n} \right)}, \label{er_brig-eq2} \\ \textstyle{{x_n}^2 P_n} &   =\textstyle{o\left( \frac{1}{\ln n} \right)}, \text{ and} \label{er_brig-eq3} \\ \textstyle{{x_n}^2 P_n} & = \textstyle{ \omega\big(\frac{(\ln n)^6}{n^2}\big)}, \label{er_brig-eq4}
 \end{align}
% $x_n  P_n =
% \Omega(n^{\epsilon})$ for a positive constant $\epsilon$, $ {x_n}=o\left( \frac{1}{n\ln n} \right)$, ${x_n}^2 P_n = o\left( \frac{1}{\ln n} \right)$
% and $ {x_n}^2 P_n
%  = \omega\big(\frac{(\ln n)^6}{n^2}\big)$,
 then
there exits some $y_n$ satisfying
\begin{align}
y_n & = \textstyle{\frac{(P_n{x_n}^2)^d}{d!}} \cdot \big[1- o\left( \frac{1}{\ln n} \right)\big]
\label{pnpb01}
\end{align}
such that Erd\H{o}s--R\'{e}nyi graph $G(n,y_n)$
  obeys
\begin{align}
 H_d(n,x_n,P_n)& \succeq_{1-o(1)} G(n,y_n) . \label{GerGb}
\end{align}

\end{lem}

We prove Lemmas \ref{brig_urig} and \ref{er_brig} in
Appendices \ref{app-prf-brig_urig} and   \ref{app-prf-er_brig}.

After establishing Lemmas \ref{brig_urig} and \ref{er_brig} to obtain Lemma \ref{lem-cpgraph-rigrig} and then using Lemma \ref{lem-cpgraph-rigrig} to get Lemma \ref{lem-cp_rig_er},
 we evaluate $z_n$ given by (\ref{ERgraph-sn-defn}) under the conditions of Theorem \ref{thm:exact_qcomposite-kcon}. First, as explained in Section \ref{sec-Confining}, to prove Theorem \ref{thm:exact_qcomposite-kcon}, we can introduce the extra condition $|\alpha_n |= o(\ln n)$. Then under the conditions of Theorem \ref{thm:exact_qcomposite-kcon} with the extra condition $| \alpha_n | = o(\ln n)$, we explain in Appendix E of the full version \cite{fullpdfaaaia} that all conditions of Lemma \ref{lem-cpgraph-rigrig}  hold, and $z_n$ given by (\ref{ERgraph-sn-defn}) satisfies
\begin{align}
\textstyle{z_n   = \frac{\ln  n + {(k-1)} \ln \ln n + {\alpha_n}-o(1)}{n}.} \label{ER-sn-kcon}
\end{align}
For $z_n$ satisfying (\ref{ER-sn-kcon}), we obtain from Lemma \ref{lem:ER:kcon} below that probability of $G(n,z_n)$ being $k$-connected
% converges to $e^{-
% \frac{e^{- \iffalse \lim\limits_ \fi \lim_{n \to \infty}{\alpha_n}}}{(k-1)!}}$ as $n\to \infty$ and hence
 can be written as $e^{-
\frac{e^{- \iffalse \lim\limits_ \fi \lim_{n \to \infty}{\alpha_n}}}{(k-1)!}} \cdot [1\pm o(1)]$, where we use $\lim_{n \to \infty}[{\alpha_n}-o(1)] = \lim_{n \to \infty}{\alpha_n}$. This result and (\ref{Gq-kcon-lower-bound-tonsb}) further induce that $\mathbb{G}_{n,d}$ under the conditions of Theorem \ref{thm:exact_qcomposite-kcon} with $| \alpha_n | = o(\ln n)$ is $k$-connected with probability at least $e^{-
\frac{e^{- \iffalse \lim\limits_ \fi \lim_{n \to \infty}{\alpha_n}}}{(k-1)!}} \times [1- o(1)]$. This proves the lower bound in Section \ref{sec-Decomposing}.

\begin{lem}[\textbf{$k$-Connectivity in an Erd\H{o}s--R\'{e}nyi graph} by {\cite[Theorem 1]{erdoskcon}}\hspace{0pt}] \label{lem:ER:kcon}
For an Erd\H{o}s--R\'enyi graph $G(n,z_n)$, if there is a sequence $\alpha_n$ with $\lim_{n \to \infty}{\alpha_n} \in [-\infty, \infty]$
such that $z_n  = \frac{\ln  n    + {(k-1)} \ln \ln n  +
 {\alpha_n}}{n}$,
%\begin{align}
%z_n  = \frac{\ln  n   +
% {\xi_n}}{n},\nonumber
%\end{align}
 then it holds that\vspace{-3pt}
 \begin{align}
 \lim_{n \to \infty}   \mathbb{P}[ G(n,z_n)\text{ is $k$-connected.} ] &  = e^{- \frac{e^{-\lim_{n \to \infty}{\alpha_n}}}{(k-1)!}} . \nonumber
 \end{align}
 \end{lem}

\subsection{More details for proving the upper bound of   Section \ref{sec-Decomposing}} \label{sec-prf-upper-bounds-details}

As explained in Section \ref{sec-prf-upper-bounds}, we prove that $\bP{\text{$\mathbb{G}_{n,d}\iffalse_{on}^{(d)}\fi$  is $k$-connected.}}$ is upper bounded by $e^{-
\frac{e^{- \iffalse \lim\limits_ \fi \lim_{n \to \infty}{\alpha_n}}}{(k-1)!}} \times [1 + o(1)]$, by first showing that $\bP{\text{$\mathbb{G}_{n,d}\iffalse_{on}^{(d)}\fi$  has a minimum degree at least $k$.}}$ is upper bounded by $e^{-
\frac{e^{- \iffalse \lim\limits_ \fi \lim_{n \to \infty}{\alpha_n}}}{(k-1)!}} \times [1 + o(1)]$. To this end, we present Lemma \ref{lem-mnd} below, where (\ref{thm-con-eq-compact-uplow}) clearly induces the desired upper bound.  In Lemma \ref{lem-mnd}, $t (K_n, P_n,d, {p_n})$ is the edge probability of $\mathbb{G}_{n,d}\iffalse_{on}^{(d)}\fi$. Note that the conditions of Lemma \ref{lem-mnd} all hold under the conditions of Theorem \ref{thm:exact_qcomposite-kcon}. In particular, from (\ref{eq_preeqnpnfngn}), $t (K_n, P_n,d, {p_n})$ in (\ref{lem-mnd-t-edgeprob}) of Lemma \ref{lem-mnd} equals the left hand side of (\ref{eq:scalinglaw-resi}) in Theorem \ref{thm:OneLaw+NodeIsolation-resi}.

\iffalse

The idea to prove the upper bound $e^{-
\frac{e^{- \iffalse \lim\limits_ \fi \lim_{n \to \infty}{\alpha_n}}}{(k-1)!}} \times [1 + o(1)]$ for $\mathbb{P}[\hspace{2pt}\mathbb{G}_{n,d}\text{ is
  $k$-connected.}\hspace{2pt}]$ has been explained in Section \ref{sec-prf-upper-bounds}. As explained, we derive the asymptotically exact probability for the property of minimum   degree being at least $k$ in the studied interest-based social network $\mathbb{G}_{n,d}$. The result is presented as Lemma \ref{lem-mnd} below,  where $t (K_n, P_n,d, {p_n})$ (i.e., $t_{n,d}$ in short) is the edge probability of $\mathbb{G}_{n,d}\iffalse_{on}^{(d)}\fi$. Note that the conditions of Lemma \ref{lem-mnd} all hold under the conditions of Theorem \ref{thm:exact_qcomposite-kcon}.

\fi

\begin{lem}[\textbf{Property of minimum degree being at least $k$ in graph $\mathbb{G}_{n,d}\iffalse_{on}^{(d)}\fi$}]\label{lem-mnd}

For a graph $\mathbb{G}_{n,d}\iffalse_{on}^{(d)}\fi$ (i.e., $\mathbb{G}_d(n, K_n,P_n, {p_n})$), if there exists a sequence $\alpha_n$ with $\lim_{n \to \infty} \alpha_n \in [-\infty, +\infty]$ such that
\begin{align}
t (K_n, P_n,d, {p_n})  & = \frac{\ln  n + (k-1) \ln \ln n   +
 {\alpha_n}}{n}  , \label{lem-mnd-t-edgeprob}
\end{align}
then it holds under $ K_n =
\omega(1)$  and $\frac{{K_n}^2}{P_n} = o(1)$ that\vspace{-3pt}
\begin{align}
&   \lim_{n \rightarrow \infty } \bP{\text{$\mathbb{G}_{n,d}\iffalse_{on}^{(d)}\fi$  has a minimum degree at least $k$.}}
%& \hspace{-57pt} \lim_{n \rightarrow \infty }  \mathbb{P} \bigg[
%\hspace{-3pt}\begin{array}{c}
%\mathbb{G}_q(n, K_n,P_n, {p_n}) \\
%\mbox{is $k$-connected.}
%\end{array}\hspace{-3pt}
%\bigg]
 \nonumber
% \\  & \hspace{-27pt}
% =   e^{- \frac{e^{-\lim_{n \to \infty}{\alpha_n}}}{(k-1)!}}  \label{thm-con-eq-compact}
\\ &    =  e^{- \frac{e^{-\lim_{n \to \infty} \alpha_{_n}}}{(k-1)!}}. \label{thm-con-eq-compact-uplow}
\end{align}
%\vspace{-5pt}
%\begin{subnumcases}
%{\hspace{-27pt} =\hspace{-2pt}} \hspace{-3pt}e^{- \frac{e^{-\alpha ^*}}{(k-1)!}},
% &\hspace{-11.5pt}\text{ if $\lim\limits_{n \to \infty}{\alpha_n}
%=\alpha ^* \in (-\infty, \infty)$,}\\  \hspace{-3pt}1, &\hspace{-11.5pt}\text{ if $\lim\limits_{n \to \infty}{\alpha_n}
%=\infty$,} \label{thm-mndx-eq-1}  \\ \hspace{-3pt} 0, &\hspace{-11.5pt}\text{ if $\lim\limits_{n \to \infty}{\alpha_n}
%=-\infty$}.\label{thm-mndx-eq-0} \label{thm-mndx-eq-e}
%\end{subnumcases}

\end{lem}

% We discuss the proof of Lemma \ref{lem-mnd} in the Appendix ???.

% =================

 We establish Lemma \ref{lem-mnd} for minimum degree in graph $\mathbb{G}_{n,d}\iffalse_{on}^{(d)}\fi$ by analyzing the asymptotically exact distribution for the number of nodes with a fixed degree, for which we present Lemma \ref{thm:exact_qcomposite2} below.

 The details of using Lemma \ref{thm:exact_qcomposite2} to prove Lemma \ref{lem-mnd} are given in Appendix F of the full version \cite{fullpdfaaaia}. We will show that to prove Lemma \ref{thm:exact_qcomposite2}, the deviation $\alpha_n$ in the lemma statement can  be confined as $\pm  o(\ln n)$. More specifically, if Lemma \ref{thm:exact_qcomposite2} holds under the extra condition $|\alpha_n |= o(\ln n)$, then Lemma \ref{thm:exact_qcomposite2} also holds regardless of the extra condition. For constant $k$ and   $|\alpha_n |= o(\ln n)$, clearly $t (K_n, P_n,d, {p_n})$ in (\ref{lem-mnd-t-edgeprob}) satisfies (\ref{peq1}).

\begin{lem}[\textbf{Possion distribution for number of nodes with a fixed degree in graph $\mathbb{G}_{n,d}\iffalse_{on}^{(d)}\fi$}]  \label{thm:exact_qcomposite2}
For graph $\mathbb{G}_{n,d}\iffalse_{on}^{(d)}\fi$ with $ K_n =
\omega(1)$  and $\frac{{K_n}^2}{P_n} = o(1)$, if
\begin{align}
t (K_n, P_n,d, {p_n}) & = \frac{\ln  n \pm o(\ln n)}{n},  \label{peq1}
\end{align}
then for a non-negative constant integer $h$, the number of nodes in
$\mathbb{G}_{n,d}\iffalse_{on}\fi$ with degree $h$ is in distribution asymptotically equivalent to a
Poisson random variable with mean $\lambda_{n,h} : = n (h!)^{-1}(n t_{n,d})^h e^{-n
t_{n,d}}$, where $t_{n,d}$ is short for $t (K_n, P_n,d, {p_n})$; i.e., as $n \to \infty$,
\begin{align}
& \hspace{-6pt}\bP{\hspace{-3pt}\begin{array}{l}\text{The number of nodes in
$\mathbb{G}_{n,d}\iffalse_{on}\fi$}\\\text{with degree $h$ equals $\ell$.}\end{array}\hspace{-3pt}}\hspace{-2pt}\Big/\hspace{-2pt}\Big[(\ell !)^{-1}{\lambda_{n,h}} ^{\ell}e^{-\lambda_{n,h}}\Big]\hspace{-2pt} \to\hspace{-2pt} 1, \nonumber \\ & \text{~~~~~~~~~~~~~~~~~~~~~~~~~~~~~~~~~~~~~~~~~~~for $\ell = 0,1, \ldots$}  \label{eq-Poisson-lemma}
\end{align}
\end{lem}

Lemma \ref{thm:exact_qcomposite2} for graph
$\mathbb{G}_{n,d}\iffalse_{on}\fi$ shows that the number of nodes
with a fixed degree follows a Poisson distribution
asymptotically. We prove Lemma \ref{thm:exact_qcomposite2} in Appendix G of the full version~\cite{fullpdfaaaia}.

\section{Conclusion} \label{sec:Conclusion}

In this paper, we investigate the resilience of an interest-based social network where users form links subject to both friendships and shared interests. To model shared interests between $n$ users, we consider that each user {independently} selects $K_n$ objects {uniformly at random} from the same pool of $P_n$ objects, and that two users establish a common-interest relation \mbox{if and only if} they share at least $d$ object(s). The network topology induced by common-interest relations is  a uniform $d$-intersection graph, denoted by $G_d(n, K_n,P_n)$. To model friendships between $n$ users, we assume that  two users are friends with probability $f_n$ so that the friendship network is represented by an Erd\H{o}s-R\'enyi graph $G(n,{f_n\iffalse_{on}\fi})$. Then an interest-based social network with both common-interest and friendship constraints becomes the intersection of a uniform $d$-intersection graph and an Erd\H{o}s-R\'enyi graph; i.e., $G_d(n, K_n,P_n)\bcap G(n,{f_n\iffalse_{on}\fi})$. To analyze the resilience of an interest-based social network  against link and node failures, we first consider that each link in $G_d(n, K_n,P_n)\bcap G(n,{f_n\iffalse_{on}\fi})$ fails independently with probability $1-g_n$ so that the remaining network after link failure is $G_d(n, K_n,P_n)\bcap G(n,{f_n\iffalse_{on}\fi})\bcap G(n,{g_n\iffalse_{on}\fi})$; i.e., we further intersect the interest-based social network $G_d(n, K_n,P_n)\bcap G(n,{f_n\iffalse_{on}\fi})$ with an Erd\H{o}s-R\'enyi graph $G(n,{g_n\iffalse_{on}\fi})$. Then we investigate connectivity of $G_d(n, K_n,P_n)\bcap G(n,{f_n\iffalse_{on}\fi})\bcap G(n,{g_n\iffalse_{on}\fi})$ even \mbox{after} an arbitrary set of $m$ {nodes fail}.
The results include the asymptotically exact probability and a
zero-one law for resilient connectivity. In addition to the formal proofs, we present experiments to confirm the results.

\appendix
%  \subsection{The Proof of Lemma \ref{lem-cp_rig_er}} \label{appsec-lem-cp_rig_er}

% \textbf{Lemma \ref{lem-cp_rig_er} (Restated).} \label{lem-cp_rig_er-restated} \textit{If $ \frac{{K_n}^2}{P_n}  =
%  o\left( \frac{1}{\ln n} \right)$, \vspace{2pt} $ \frac{K_n}{P_n} = o\left( \frac{1}{n\ln n} \right)$, \vspace{1pt} $ K_n =
% \Omega(n^{\epsilon})$ for a positive constant $\epsilon$, and $\frac{{K_n}^2}{P_n} = \omega\big(\frac{(\ln n)^6}{n^2}\big)$, then there exists a sequence $z_n$ satisfying
% \begin{align}
% z_n = t_{n,d} \times \textstyle{\big[1-o\big(\frac{1}{\ln n}\big)\big]} \label{ERgraph-sn-defn-restated}
% \end{align}
%  such that
% graph $\mathbb{G}_{n,d}\iffalse_{on}^{(d)}\fi$ contains an Erd\H{o}s--R\'{e}nyi graph $G(n,z_n)$ as a spanning subgraph with probability $1-o(1)$ (when we couple the two graphs on the same probability space and define them on the same node set), where $t_{n,d}$ is the edge probability of $\mathbb{G}_{n,d}\iffalse_{on}^{(d)}\fi$.}

% \textbf{Notation for coupling between random graphs:}

% We present the proofs of Lemmas \ref{er_brig} and \ref{er_brig-weaker} together in
% Appendix \ref{app-prf-er_brig}.

% \subsection{An asymptotic expression of the edge probability $s_{n,d} $ of uniform $d$-intersection graph $G_d(n, K_n,P_n)$}

% \begin{lem} \label{lem_eval_psq}
% If $K_n = \omega(\ln n)$ and $\frac{{K_n}^2}{P_n} = o\big(\frac{1}{\ln n}\big)$, then\\$s_{n,d}
% = \frac{1}{d!} \big( \frac{{K_n}^2}{P_n} \big)^{d} \times [1\pm o\big(\frac{1}{\ln n}\big)]$.
% \end{lem}

\subsection{{Proof of Lemma \ref{lem-cpgraph-rigrig} Using
Lemmas \ref{brig_urig} and \ref{er_brig}}} \label{app-prf-lem-cpgraph-rigrig}

We complete the proof of Lemma \ref{lem-cpgraph-rigrig} by using
Lemmas \ref{brig_urig} and \ref{er_brig}. Lemmas \ref{brig_urig} and \ref{er_brig} have been presented on Page \pageref{er_brig}, and will be proved in
Appendices \ref{app-prf-brig_urig} and   \ref{app-prf-er_brig}, respectively.
\iffalse

{ \normalsize
\begin{table}[!h]
\normalsize \caption{\normalsize Different Graphs.}
\begin{center}  %\lable{table:notation}
 \begin{tabular}
 {!{\vrule width 1pt}l|l|l!{\vrule width 1pt}}
   \Xhline{1\arrayrulewidth}
    \hspace{-2pt}Notation\hspace{-3pt} & \hspace{-2pt}Kind\hspace{-3pt} & \hspace{-2pt}Edge prob.\hspace{-3pt}  \\ \Xhline{1\arrayrulewidth}
   \hspace{-2pt}$G_d(n,K_n,P_n)$\hspace{-3pt} & \hspace{-2pt}uniform random intersection\hspace{-3pt} & \hspace{-2pt}$p_d$\hspace{-3pt} \\ \hline
   \hspace{-2pt}$H_d(n,x_n,P_n)$\hspace{-3pt} & \hspace{-2pt}binomial random intersection\hspace{-3pt} & \hspace{-2pt}$p_b$\hspace{-3pt}  \\ \hline
   \hspace{-2pt}$G(n,y_n)$\hspace{-3pt} & \hspace{-2pt}Erd\H{o}s--R\'{e}nyi\hspace{-3pt} & \hspace{-2pt}$y_n$\hspace{-3pt} \\
   \Xhline{1\arrayrulewidth}
 \end{tabular}
 \end{center}
\end{table}
}

\fi
 We first explain that given the conditions of Lemma \ref{lem-cpgraph-rigrig}:
 \begin{align}
 \textstyle{\frac{{K_n}^2}{P_n} } & = \textstyle{ o\left( \frac{1}{\ln n} \right)}, \label{pnKn2-toneq1} \\  \textstyle{\frac{K_n}{P_n}} & = \textstyle{o\left( \frac{1}{n\ln n} \right)}, \label{pnKn2-toneq2} \\ \textstyle{ K_n } & = \textstyle{\Omega(n^{\epsilon})}\text{ for a positive constant $\epsilon$}, \label{pnKn2-toneq3} \\ \textstyle{\frac{{K_n}^2}{P_n}} & = \textstyle{\omega\big(\frac{(\ln n)^6}{n^2}\big)}, \label{pnKn2-toneq4}
 \end{align}
 all
conditions in Lemmas \ref{brig_urig} and \ref{er_brig} are true;
i.e.,
 \begin{align}
\textstyle{ K_n } & = \omega(\ln n), \label{pnKn2-toneqa1} \\  \textstyle{ \frac{{K_n}^2}{P_n} } & = \textstyle{o\left( 1 \right)}, \label{pnKn2-toneqa2} \\ \textstyle{{x_n}} &   =\textstyle{o\left( \frac{1}{n\ln n} \right)}, \label{pnKn2-toneqa3} \\ \textstyle{{x_n}^2 P_n} & = \textstyle{ o\left( \frac{1}{\ln n} \right)}, \text{ and} \label{pnKn2-toneqa4}  \\ \textstyle{ {x_n}^2 P_n} & = \textstyle{\omega\big(\frac{(\ln n)^6}{n^2}\big)}, \label{pnKn2-toneqa5}
 \end{align}
all hold, where $x_n$ is defined in (\ref{pnKn}).

Clearly, (\ref{pnKn2-toneq3}) implies (\ref{pnKn2-toneqa1}) given $\lim_{n \to \infty} {n^{\epsilon}}/({\ln n}) = \infty$. Also, (\ref{pnKn2-toneqa2}) implies (\ref{pnKn2-toneq1}).
Using (\ref{pnKn2-toneq3}) in (\ref{pnKn}), \f
\begin{align}
 x_n & = \textstyle{\frac{K_n}{P_n}  \cdot \Big[1 - O\Big(\sqrt{\frac{3\ln
n}{n^{\epsilon}}}\Big)\hspace{1pt}\Big] } \label{pnKn2-tonabc} \\  & = \textstyle{\frac{K_n}{P_n}  \cdot [1 - o(1)]} \label{pnKn2}.
 \end{align}
Then we obtain the following. First, (\ref{pnKn2}) and (\ref{pnKn2-toneq2}) together yield (\ref{pnKn2-toneqa3}).
Second, (\ref{pnKn2}) and (\ref{pnKn2-toneq1}) induce
(\ref{pnKn2-toneqa4}). Third,
(\ref{pnKn2}) and (\ref{pnKn2-toneq4}) lead
to (\ref{pnKn2-toneqa5}). Therefore, all conditions in Lemmas \ref{brig_urig}
and \ref{er_brig} hold.

We use $y_n$ defined in (\ref{pnpb01}). By \cite[Fact
3]{2013arXiv1301.0466R} on the transitivity of graph coupling, we
use (\ref{eq_brig_urig}) in Lemma \ref{brig_urig} and (\ref{GerGb})
in Lemma \ref{er_brig} to obtain
\begin{align}
G_d(n,K_n,P_n) & \succeq_{1-o(1)}G(n,y_n) . \label{GerGurig}
\end{align}
From (\ref{pnpb01}) and (\ref{pnKn2-tonabc}), we derive
\begin{align}y_n \hspace{-2pt} =\hspace{-2pt}
\textstyle{\frac{1}{d!}  \hspace{-2pt}\cdot\hspace{-2pt} \frac{{K_n}^{2d}}{{P_n} ^{d}}} \hspace{-2pt}\cdot\hspace{-2pt}
 \Big[1 - O\Big(\sqrt{\frac{3\ln
n}{n^{\epsilon}}}\Big)\hspace{1pt}\Big]^{2d} \hspace{-2pt}=\hspace{-2pt}
\textstyle{\frac{1}{d!}  \hspace{-2pt}\cdot\hspace{-2pt} \frac{{K_n}^{2d}}{{P_n} ^{d}}} \hspace{-2pt}\cdot\hspace{-2pt}
\big[1- o\left( \frac{1}{\ln n} \right)\big]. \label{GerGurigsba}
\end{align}

Given $K_n = \omega(\ln n)$ (which clearly holds from the condition $ K_n =
\Omega(n^{\epsilon})$ for a positive constant $\epsilon$) and the condition $\frac{{K_n}^2}{P_n} = o\big(\frac{1}{\ln n}\big)$, we use Lemma \ref{lem_eval_psq}-Property (ii) to obtain \begin{align}\textstyle{s_{n,d}
= \frac{1}{d!} \big( \frac{{K_n}^2}{P_n} \big)^{d} \times \big[1\pm o\big(\frac{1}{\ln n}\big)\big]. }\label{GerGurigsbb}
\end{align}
Summarizing (\ref{GerGurigsba}) and (\ref{GerGurigsbb}), we obtain $y_n = s_{n,d} \times \big[1\pm o\big(\frac{1}{\ln n}\big)\big]$; i.e., (\ref{ERgraph-sn-defn-reduced}) of Lemma \ref{lem-cpgraph-rigrig}  is proved. From \cite[Fact 3]{zz}, for Erd\H{o}s--R\'enyi graphs
$G(n,y_n')$ and $G(n,y_n'')$, if $y_n ' \geq y_n''$, then
$G(n,y_n')\succeq G(n,y_n'')$. Hence, we can replace ``$\pm$'' in the above expression of $y_n$ by ``$-$''; i.e., we can set $y_n = s_{n,d} \times  \big[1- o\big(\frac{1}{\ln n}\big)\big]$ to ensure (\ref{GerGurig}), which means that graph $G_d(n, K_n, P_n)$ contains an Erd\H{o}s--R\'{e}nyi graph $G(n,y_n)$ as a spanning subgraph with probability $1-o(1)$ (when we couple the two graphs on the same probability space and define them on the same node set). Then the proof of Lemma \ref{lem-cpgraph-rigrig} is
completed. \qeda

In the above proof of Lemma \ref{lem-cpgraph-rigrig}, we have used Lemma \ref{lem_eval_psq}, which provides asymptotic expressions of the edge probability $s_{n,d} $ of uniform $d$-intersection graph $G_d(n, K_n,P_n)$. Lemma \ref{lem_eval_psq} is proved in Appendix H of the full version \cite{fullpdfaaaia}.

% $s_{n,d} $ denotes the edge probability of uniform $d$-intersection graph $G_d(n, K_n,P_n)$ (i.e., the probability that two  nodes share at least $d$ objects) ??

\begin{lem} \label{lem_eval_psq}
The following two properties hold, where $s_{n,d} $ denotes the edge probability of uniform $d$-intersection graph $G_d(n, K_n,P_n)$:
\begin{itemize}
\item[(i)] If $K_n = \omega(1)$ and $\frac{{K_n}^2}{P_n} = o(1)$, then\\$s_{n,d}
= \frac{1}{d!} \big( \frac{{K_n}^2}{P_n} \big)^{d} \times [1\pm o(1)]$; i.e., $s_{n,d}
\sim \frac{1}{d!} \big( \frac{{K_n}^2}{P_n} \big)^{d}$.\vspace{3pt}
\item[(ii)] If $K_n = \omega(\ln n)$ and $\frac{{K_n}^2}{P_n} = o\big(\frac{1}{\ln n}\big)$, then\\$s_{n,d}
= \frac{1}{d!} \big( \frac{{K_n}^2}{P_n} \big)^{d} \times [1\pm o\big(\frac{1}{\ln n}\big)]$.
\end{itemize}
\end{lem}

\subsection{{Proof of Lemma \ref{brig_urig}}}\label{app-prf-brig_urig}

By \cite[Lemma 4]{Rybarczyk}, if $x_n P_n = \omega\left( \ln n
\right)$, and for all $n$ sufficiently large,
\begin{align}
K_{n}  & \geq x_n P_n + \sqrt{3(x_n P_n + \ln n) \ln n}  ,
\label{Knbig}
\end{align}
then
\begin{align}
 G_d(n,K_n,P_n) & \succeq_{1-o(1)}H_d(n,x_n,P_n) .
\end{align}

Therefore, the proof of Lemma \ref{brig_urig} is completed once we
show $x_n P_n = \omega\left( \ln n \right)$ and (\ref{Knbig}) with
$x_n$ defined in (\ref{pnKn}). From conditions\vspace{1pt} $K_n =
\omega\left( \ln n \right)$ and $x_n = \frac{K_n}{P_n}
 \left(1 - \sqrt{\frac{3\ln
n}{K_n }}\hspace{2pt}\right)$, we first obtain $x_n P_n  =
\omega\left( \ln n \right)$ and then
 for all $n$ sufficiently large,
\begin{align}
&  K_n - \left[ x_n P_n + \sqrt{3(x_n P_n + \ln n) \ln n}
\hspace{1.5pt}\right] \nonumber \\ & = K_n \sqrt{\frac{3\ln n}{K_n
}} - \sqrt{3\left[ K_n \left(1 - \sqrt{\frac{3\ln n}{K_n
}}\hspace{2pt}\right) + \ln n\right] \ln n} \nonumber
\\  & = \sqrt{3K_n\ln n}  -
\sqrt{3\left[K_n  \hspace{-1pt}+ \hspace{-1pt} \sqrt{\ln n} \left(
\sqrt{\ln n} \hspace{-1pt}- \hspace{-1pt}
\sqrt{3K_n}\hspace{2pt}\right) \right ] \hspace{-1pt} \ln n}
\nonumber \\  & \geq \sqrt{3K_n\ln n} - \sqrt{3K_n\ln n} \nonumber
\\  & =  0,
\end{align}
where we use $K_n \geq \ln n$ for all $n$ sufficiently large (this
holds from condition $K_n = \omega\left( \ln n \right)$). Then it is
clear that Lemma \ref{brig_urig} is proved. \qeda

\subsection{{Proof of Lemma \ref{er_brig}}} \label{app-prf-er_brig}

%binomial random intersection graph $G_d(n,P_n,x_n)$
%The following argument on graph coupling for proving Lemma
%\ref{er_brig} is motivated by Rybarczyk's work \cite{zz}.
% Graph $H(n,X)$ defined on node set $V$ have edge
%set constructed by sampling  When $X$ equals $0$, graph $H(n,X)$ is
%empty.
We number the objects in the object pool of size $P_n$ by $1,2,\ldots,P_n$.
In binomial $d$-intersection graph $H_d(n,P_n,x_n)$, let $\mathcal
{U}_i$ be the set of sensors assigned with object $\kappa_i$
($i=1,2,\ldots,P_n$). Then $U_i$ denoting the cardinality of $\mathcal
{U}_i$ (i.e., $U_i:=|\mathcal {U}_i|$) obeys a binomial distribution
$\textrm{Bin}(n, x_n)$, with $n$ as the number of trials, and $x_n$
as the success probability in each trial. Clearly, we can generate
the random set $\mathcal {U}_i$ in the following equivalent manner:
First draw the cardinality $U_i$ from the distribution
$\textrm{Bin}(n, x_n)$, and then choose $U_i$ distinct nodes
uniformly at random from the set $\mathcal{V}_n$ of all $n$ nodes ($\mathcal{V}_n=\{v_1,v_2,\ldots,v_n\}$).

Given  $\mathcal {U}_i$ defined above, we generate a graph $H(\mathcal{U}_i)$ on node set $\mathcal{V}_n$ as follows. We construct
the graph $H(\mathcal{U}_i)$ by establishing edges between any and only pair
of nodes in $\mathcal {U}_i$; i.e., $H(\mathcal{U}_i)$ has a clique on
$\mathcal {U}_i$ and no edges between nodes outside of this clique.
If a given realization of the random variable $U_i$ satisfies $U_i <
2$, then the corresponding instantiation of $H(\mathcal{U}_i)$ will be an
empty graph.

We now explain the connection between $H(\mathcal{U}_i)$ and the binomial
  $d$-intersection graph $H_d(n,P_n,x_n)$.
We let an operator ${\cal{O}}_d$ take a multigraph
\cite{Citeulike:505396} with possibly multiple edges between two
nodes as its argument. The operator returns a simple graph with an
undirected edge between two nodes $i$ and $j$, if and only if the
input multigraph has at least $d$ edges between these nodes. Recall
that two nodes in $H_d(n,P_n,x_n)$
  need to share at least $d$ objects to
have an edge in between. Then, with $H(\mathcal{U}_1), \ldots, H(\mathcal{U}_{P_n})$
generated independently, it is straightforward to see
\begin{equation}
{\cal{O}}_d \left(\bigcup_{i=1}^{P_n}  H(\mathcal{U}_i) \right)
=_{\textrm{st}}  H_d(n,P_n,x_n), \label{eq:osy_new_1}
\end{equation}
with $=_{\textrm{st}}$ denoting statistical equivalence.
% two nodes
%establish an edge in $H_d(n,P_n,x_n)$ if and only if there exist at
%least $d$ different number of $i$'s such that the two nodes have an
%edge in each of these $H(\mathcal{U}_i)$.

We will introduce auxiliary random graphs $L(n,B)$ and $L_d(n,B)$,
both defined on the $n$-size node set $\mathcal{V}_n = \{v_1,v_2,\ldots,v_n\}$, where $B$ is a random integer variable. The motivation for defining $L(n,B)$ and $L_d(n,B)$ is that they serve as an intermediate step to build the connection between the above binomial
  $d$-intersection graph $H_d(n,P_n,x_n)$ and an Erd\H{o}s--R\'enyi graph. More specifically,
\begin{itemize}
\item on the one hand, given $U_i$ defined above, we build the connection between $L(n,\big\lfloor U_i/2\big\rfloor)$ and $H(\mathcal{U}_i)$, in order to find the relationship between $L_d\big(n,\sum_{i=1}^{P_n}\big\lfloor U_i/2\big\rfloor \big)$ and the binomial
  $d$-intersection graph $H_d(n,P_n,x_n)$;
\item on the other hand, when $Z$ is a Poisson random variable, $L(n,Z)$ becomes an
Erd\H{o}s--R\'{e}nyi graph;
\item given the above two points, we further find  the relationship between $L_d\big(n,\sum_{i=1}^{P_n}\big\lfloor U_i/2\big\rfloor \big)$ and $L(n,Z)$ for a Poisson random variable $Z$. Then summarizing all points, we build the connection between the binomial
  $d$-intersection graph $H_d(n,P_n,x_n)$ and an Erd\H{o}s--R\'enyi graph.
\end{itemize}
We now define $L(n,B)$ and $L_d(n,B)$ on the node set $\mathcal{V}_n = \{v_1,v_2,\ldots,v_n\}$ for a random integer variable $B$. For different nodes $v_i$ and $v_j$, we use $\text{edge}(v_i, v_j)$ to denote an undirected edge between nodes $v_i$ and $v_j$ so there is no difference between $\text{edge}(v_i, v_j)$ and $\text{edge}(v_j, v_i)$. For the $n$ nodes in $\mathcal{V}_n = \{v_1,v_2,\ldots,v_n\}$, the number of possible edges is $\binom{n}{2}$ (i.e., the number of ways to select two unordered nodes from $n$ nodes). Among these $\binom{n}{2}$ edges, we select one edge uniformly at random at each time. We repeat the selection $b$ times independently for an integer $b$. Note that at each time, an edge is selected from the $\binom{n}{2}$ edges, so we have that even if an edge has already been selected, it may get selected again next time. In other words, the selections are done \emph{with repetition} since it is possible that an edge gets selected multiple times. After the $b$ times of selection, we obtain $b$ edges where several edges may be the same. These $b$ edges constitute
a multiset $\mathcal{M}(b)$, where a multiset is a generalization of a set such that  unlike a set, a multiset allows multiple elements to take the same value. Given an integer $b$, after obtaining a  multiset $\mathcal{M}(b)$ according to the above procedure, we now construct graphs $L(n,b)$ and $L_d(n,b)$, which are both defined on the node set $\mathcal{V}_n = \{v_1,v_2,\ldots,v_n\}$. An edge is put in graph $L(n,b)$ if and only if it appears at least once in the multiset $\mathcal{M}(b)$, while an edge is put in graph $L_d(n,b)$ if and only if it appears at least $d$ times in the multiset $\mathcal{M}(b)$. Now given graphs $L(n,b)$ and $L_d(n,b)$ for an integer $b$, we define graphs $L(n,B)$ and $L_d(n,B)$ for an integer-valued random variable $B$ as follows: we let $L(n,B)$ be $L(n,b)$ with probability $\mathbb{P}[B=b]$, and let $L_d(n,B)$ be $L_d(n,b)$ with probability $\mathbb{P}[B=b]$.

\iffalse

We only take the distinct edges and let the number of distinct edges be $y$; i.e., if an edge has been selected for $z$ times, we only keep one copy and delete the rest $(z-1)$ copies. Now we have $y$ distinct edges, where each edge is between two nodes. We use these $y$ edges to define

where $B$ is a
non-negative random integer variable.

We sample $B$ node pairs
\emph{with repetition} from all pairs of $\mathcal{V}_n$ (a pair is
unordered). In graph $L(n,B)$ (resp., $L_d(n,B)$), two nodes have an
edge in between if and only if the node pair is sampled at least
once (resp., $d$ times).

Note that $B$ can also be a
fixed value with probability $1$.

\fi

With $H(\mathcal{U}_i)$ and $L(n,B)$ given above, we show a coupling below
under which random graph $L(n,\big\lfloor U_i/2\big\rfloor)$ is a
subgraph of random graph $H(\mathcal{U}_i)$; i.e.,
\begin{align}
 H(\mathcal{U}_i) & \succeq L(n,\big\lfloor U_i/2\big\rfloor) . \label{cpHG}
\end{align}
By definition, graph $L(n,\big\lfloor U_i/2\big\rfloor)$ has at most
$\big\lfloor U_i/2\big\rfloor$ edges and thus contains non-isolated
nodes with a number (denoted by $\ell$) at most $2\cdot \big\lfloor
U_i/2\big\rfloor \leq U_i$, where a node is non-isolated if it has a link with at least another node, and a node is isolated if it has no link with any other node. Given an instance $\mathcal {L}$ of
random graph $L(n,\big\lfloor U_i/2\big\rfloor)$, we construct set
$\mathcal {U}_i$ as the union of the $\ell$ number non-isolated
nodes in $\mathcal {L}$ and the rest $(U_i -\ell)$ nodes selected
uniformly at random from the rest $(n-\ell)$ isolated nodes in
$\mathcal {L}$. Since graph $H(\mathcal{U}_i)$ contains a clique of $\mathcal
{U}_i$, it is clear that the induced instance of $H(\mathcal{U}_i)$ is a
supergraph of the instance $\mathcal {L}$ of graph $L(n,\big\lfloor
U_i/2\big\rfloor)$. Then the proof of (\ref{cpHG}) is completed.

Now based on $L(n,\big\lfloor U_i/2\big\rfloor)$, we construct a
graph defined on node set $\mathcal{V}_n$. We add an edge between two
nodes in this graph if and only if there exist at least $d$
different number of $i$ such that the two nodes have an edge in each
of these $L(n,\big\lfloor U_i/2\big\rfloor)$. By the independence of
$U_i$ ($i=1,2,\ldots,P_n$) and the definition of $L_d(n,B)$ above,
it is clear that such induced graph is statistically equivalent to
$L_d\big(n,\sum_{i=1}^{P_n}\big\lfloor U_i/2\big\rfloor \big)$.
Namely, we have
\begin{align}
{\cal{O}}_d \left(\bigcup_{i=1}^{P_n}  L(n,\big\lfloor
U_i/2\big\rfloor) \right) =_{\textrm{st}}
L_d\big(n,\sum_{i=1}^{P_n}\big\lfloor U_i/2\big\rfloor \big),
\label{eq:osy_new_2}
\end{align}
where ``$=_{\textrm{st}}$'' means statistical  equivalence.

In view of (\ref{eq:osy_new_1}), (\ref{cpHG}), and
(\ref{eq:osy_new_2}), we see
\begin{align}
 H_d(n,P_n,x_n) & \succeq L_d(n, Y), \label{HqnY}
\end{align}
where $Y$ is defined via
\begin{align}
 Y: = \sum_{i=1}^{P_n}W_i,\label{newzY}
\end{align}
with
\begin{align}
W_i : = \big\lfloor U_i/2\big\rfloor = \m{\frac{1}{2}}(U_i-
\textrm{I}_{[U_i\textrm{ is odd}]}). \label{eqWi}
\end{align}

We now explore a bound of $Y$ based on (\ref{newzY}) and
(\ref{eqWi}). For a random variable $\mathcal {R}$, we denote its
expected value (i.e., mean) and variance by $\mathbb{E}[\mathcal
{R}]$ and $\textrm{Var}[\mathcal {R}]$, respectively. As noted,
$U_i$ obeys a binomial distribution $\textrm{Bin}(n, x_n)$. Then
\begin{align}
\mathbb{E}[U_i] & = \sum_{a=0,1,\ldots, n} \Bigg[ a \cdot
\binom{n}{a}{x_n}^a(1-x_n)^{n-a}\Bigg]  \nonumber
\\  & = nx_n \sum_{a=0,1,\ldots, n} \Bigg[
\binom{n-1}{a-1}{x_n}^{a-1}(1-x_n)^{n-a}\Bigg]  \nonumber
\\  & = nx_n [x_n+(1-x_n)]^{n-1} \nonumber
\\  & =  nx_n,  \label{evi1}
\end{align}
and
\begin{align}
 &\mathbb{E}\big[ \textrm{I}_{[U_i\textrm{ is odd}]}\big]   \nonumber
\\  & \quad = \mathbb{P}[U_i\textrm{ is odd}]  \nonumber
\\  & \quad = \sum_{a=1,3,\ldots, n -
\textrm{I}_{[U_i\textrm{ is even}]}}
\binom{n}{a}{x_n}^a(1-x_n)^{n-a} \nonumber
\\  & \quad = \frac{1}{2} \sum_{a=0,1,\ldots, n}
\binom{n}{a}{x_n}^a(1-x_n)^{n-a}  \nonumber
\\  &  \quad \quad  - \frac{1}{2} \sum_{a=0,1,\ldots,
n} \binom{n}{a}(-x_n)^{a}(1-x_n)^{n-a}  \nonumber
\\  & \quad = \m{\frac{1}{2}} [x_n + (1-x_n)]^{n} - \m{\frac{1}{2}} [-x_n +
(1-x_n)]^{n}  \nonumber
\\  & \quad = \m{\frac{1}{2}}[1 - (1-2x_n)^{n}] .  \label{evi2}
\end{align}
Applying (\ref{evi1}) and (\ref{evi2}) to (\ref{eqWi}), and using
$ {x_n}=o\big(\frac{1}{n\ln n}\big)$ (i.e., (\ref{er_brig-eq2}) of Lemma \ref{er_brig}), we derive
\begin{align}
& \mathbb{E}[W_i] \nonumber
\\   &  = \m{\frac{1}{2}}\mathbb{E}[U_i] -
\m{\frac{1}{2}}\mathbb{E}\big[ \textrm{I}_{[U_i\textrm{ is
odd}]}\big]  \label{Ewi}
\\  & = \m{\frac{1}{2}}nx_n  - \m{\frac{1}{4}} + \m{\frac{1}{4}}(1-2x_n)^{n}   \nonumber
\\  & = \m{\frac{1}{2}}nx_n  - \m{\frac{1}{4}} + \m{\frac{1}{4}}\big[1-2nx_n
+ 2n(n-1){x_n}^2 \pm O\big(n^3{x_n}^3\big)\big]  \nonumber
\\  & = \m{\frac{1}{2}} n(n-1){x_n}^2 \pm O\big(n^3{x_n}^3\big) \nonumber
\\  & =  \m{\frac{1}{2}} n(n-1){x_n}^2 \cdot [1 \pm o(n{x_n})].
\label{exWi-stronger}
% \\  & =  \m{\frac{1}{2}} n(n-1){x_n}^2 \cdot [1 \pm o(1)].
% \label{exWi}
\end{align}

%\begin{align}
% &\mathbb{E}\big[ \textrm{I}_{[U_i\textrm{ is odd}]}\big]   \nonumber
%\\  & \quad = \mathbb{P}[U_i\textrm{ is odd}]  \nonumber
%\\  & \quad = \sum_{a=1,3,\ldots, n -
%\textrm{I}_{[U_i\textrm{ is even}]}}
%\binom{n}{a}{x_n}^a(1-x_n)^{n-a} \nonumber
%\\  & \quad = \frac{1}{2} \sum_{a=0,1,\ldots, n}
%\binom{n}{a}{x_n}^a(1-x_n)^{n-a} \nonumber
%\\  & \quad \quad - \frac{1}{2} \sum_{a=0,1,\ldots,
%n} \binom{n}{a}(-x_n)^{a}(1-x_n)^{n-a}  \nonumber
%\\  & \quad = \m{\frac{1}{2}} [x_n + (1-x_n)]^{n} - \m{\frac{1}{2}} [-x_n +
%(1-x_n)]^{n}  \nonumber
%\\  & \quad = \m{\frac{1}{2}}[1 - (1-2x_n)^{n}] . \label{evi2}
%\end{align}

From (\ref{eqWi}), it holds that
\begin{align}
& \textrm{Var}[2W_i]  \nonumber
\\ & = \textrm{Var}\big[U_i- \textrm{I}_{[U_i\textrm{
is odd}]}\big]  \nonumber
\\  &= \textrm{Var}[U_i]+ \textrm{Var}[\textrm{I}_{[U_i\textrm{ is
odd}]}]  - 2 \textrm{Cov}[U_i, \textrm{I}_{[U_i\textrm{ is odd}]}],
\label{var2wi}
\end{align}
where $\textrm{Cov}[U_i, \textrm{I}_{[U_i\textrm{ is odd}]}]$
denoting the covariance between $U_i$ and $\textrm{I}_{[U_i\textrm{
is odd}]}$ is given by
\begin{align}
& \textrm{Cov}[U_i, \textrm{I}_{[U_i\textrm{ is odd}]}]  \nonumber
\\ & = \mathbb{E}\big[(U_i-\mathbb{E}[U_i])\big(\textrm{I}_{[U_i\textrm{ is odd}]}
- \mathbb{E}[\textrm{I}_{[U_i\textrm{ is odd}]}]\big)\big] \nonumber
\\ & = \mathbb{E}[U_i\textrm{I}_{[U_i\textrm{ is odd}]}] - \mathbb{E}[U_i]\mathbb{E}[\textrm{I}_{[U_i\textrm{ is
odd}]}].
   \label{var2wi1}
\end{align}
Clearly, it holds that $U_i\textrm{I}_{[U_i\textrm{ is odd}]} \geq
\textrm{I}_{[U_i\textrm{ is odd}]}$, inducing
\begin{align}
\mathbb{E}[U_i\textrm{I}_{[U_i\textrm{ is odd}]}] & \geq
\mathbb{E}[\textrm{I}_{[U_i\textrm{ is odd}]}].
   \label{var2wi1ViIVi}
\end{align}
From (\ref{evi1}) and (\ref{evi2}), we further obtain
\begin{align}
&\mathbb{E}[U_i]\mathbb{E}[\textrm{I}_{[U_i\textrm{ is odd}]}] -
\mbox{$\frac{3}{2}$} \cdot (\mathbb{E}[U_i] -
\mathbb{E}[\textrm{I}_{[U_i\textrm{ is odd}]}]) \nonumber
\\ & = n x_n \cdot \mbox{$\frac{1}{2}$}[1 - (1-2x_n)^{n}]  - \mbox{$\frac{3}{2}$} \big\{ n x_n - \mbox{$\frac{1}{2}$}[1 -
(1-2x_n)^{n}]\big\}  \nonumber
\\ & = -nx_n + \m{\frac{3}{4}} -
(\m{\frac{1}{2}}nx_n+\m{\frac{3}{4}})(1-2x_n)^{n}  \nonumber
\\ & \leq -nx_n + \m{\frac{3}{4}} -
(\m{\frac{1}{2}}nx_n+\m{\frac{3}{4}})(1-2nx_n+\m{\frac{4}{3}}n^2{x_n}^2)
\nonumber
\\ & = - \m{\frac{2}{3}} n^2{x_n}^2 \leq 0,  \label{var2wi2}
\end{align}
where the step involving the first ``$\leq$'' uses the inequality
$(1-2x_n)^{n} \geq 1-2nx_n+\m{\frac{4}{3}}n^2{x_n}^2$ for all $n$
sufficiently large, which is derived from a Taylor expansion of the
binomial series $(1-2x_n)^{n}$, given $ {x_n}=o\big(\frac{1}{n\ln n}\big)$ (i.e., (\ref{er_brig-eq2}) of Lemma \ref{er_brig}).

Using (\ref{var2wi1ViIVi}) and (\ref{var2wi2}) in (\ref{var2wi1}),
it follows that
\begin{align}
 \textrm{Cov}[U_i, \textrm{I}_{[U_i\textrm{ is odd}]}]  & \geq \mbox{$\frac{5}{2}$} \mathbb{E}[\textrm{I}_{[U_i\textrm{ is odd}]}] - \mbox{$\frac{3}{2}$} \mathbb{E}[U_i].  \label{var2wi3}
\end{align}
For binomial random variable $U_i$ and Bernoulli random variable
$\textrm{I}_{[U_i\textrm{ is odd}]}$, it is clear that
\begin{align}
\textrm{Var}[U_i] & \leq \mathbb{E}[U_i], \label{varvi}
\end{align}
and
\begin{align}
\textrm{Var}[\textrm{I}_{[U_i\textrm{ is odd}]}] & \leq
\mathbb{E}[\textrm{I}_{[U_i\textrm{ is odd}]}]. \label{varviI}
\end{align}

Applying (\ref{var2wi3}) (\ref{varvi}) and (\ref{varviI}) to
(\ref{var2wi}), we have
\begin{align}
 \textrm{Var}[2W_i] & \leq \mathbb{E}[U_i] +
\mathbb{E}[\textrm{I}_{[U_i\textrm{ is odd}]}] \nonumber
\\ & \quad  - 5\mathbb{E}[\textrm{I}_{[U_i\textrm{ is odd}]}] + 3\mathbb{E}[U_i]  \nonumber
\\ & = 4 (\mathbb{E}[U_i] -
\mathbb{E}[\textrm{I}_{[U_i\textrm{ is odd}]}]),
%\\ & = 8 \mathbb{E}[W_i], \nonumber
\end{align}
which along with (\ref{Ewi}) yields $\textrm{Var}[2W_i] \leq 8
\mathbb{E}[W_i]$; i.e.,
\begin{align}
 \textrm{Var}[W_i]  & \leq  2 \mathbb{E}[W_i]. \label{varWi}
\end{align}

Considering the independence of $W_i$ ($i=1,2,\ldots,P_n$), for $Y =
\sum_{i=1}^{P_n}W_i$ given in (\ref{Y}), we use (\ref{varWi}) to
derive
\begin{align}
 \textrm{Var}[Y]  & \leq  2 \mathbb{E}[Y]. \label{varYEY}
\end{align}

From $Y = \sum_{i=1}^{P_n}W_i$, (\ref{exWi-stronger}), and the fact that $\mathbb{E}[W_i]$ for each $i$ is the same, we obtain
\begin{align}
\mathbb{E}[Y]  & =   \m{\frac{1}{2}} n(n-1)P_n {x_n}^2 \cdot [1 \pm o(n{x_n})]
 \label{varYEY2-stronger-sbsbabcd}.
\end{align}
Note that Lemma \ref{er_brig} has conditions (\ref{er_brig-eq2}) and (\ref{er_brig-eq4}) (i.e., ${x_n} = o\left( \frac{1}{n\ln n} \right)$ and ${x_n}^2 P_n = \omega\big(\frac{(\ln n)^6}{n^2}\big)$). Using these in (\ref{varYEY2-stronger-sbsbabcd}), we have
% From (\ref{exWi-stronger}), $Y = \sum_{i=1}^{P_n}W_i$, and condition
% $n^2{x_n}^2 P_n
%  = \omega(1)$, \h
% \begin{align}
% \mathbb{E}[Y]  & =   \m{\frac{1}{2}} n(n-1)P_n {x_n}^2 \cdot [1 \pm o(n{x_n})]
% %\label{varYEY2-stronger}
% \\  & =  \m{\frac{1}{2}} n(n-1)P_n {x_n}^2 \cdot [1 \pm
% o(1)] = \omega(1) .
% %\label{varYEY2}
% \end{align}
% \begin{subnumcases}{\mathbb{E}[Y]\hspace{-2pt}  =\hspace{-2pt} }
% \hspace{-5pt}\omega\big((\ln n)^6\big),&\text{ for Lemma \ref{er_brig}} . \label{GerGb-ton-tac1} \\ \hspace{-5pt}\omega(1),&\text{ for Lemma \ref{er_brig-weaker}} \label{GerGb-ton-tac2}
% \end{subnumcases}
\begin{align}
\mathbb{E}[Y]  =\textstyle{\frac{1}{2}n(n-1)P_n {x_n}^2 \cdot \big[1 \pm o\big(\frac{1}{\ln n}\big)\big]}\label{GerGb-ton-tac4}
\end{align}
and
\begin{align}
\mathbb{E}[Y]  =
\textstyle{\omega\big((\ln n)^6\big)}\label{GerGb-ton-tac2}.
\end{align}

% \begin{subnumcases}{\hspace{-30pt}\mathbb{E}[Y]  \hspace{-2pt}= \hspace{-2pt}}
% \hspace{-5pt}\textstyle{\frac{1}{2}n(n-1)P_n {x_n}^2 \cdot \big[1 \pm o\big(\frac{1}{\ln n}\big)\big]}\hspace{-.5pt},&\text{\hspace{-17pt}for Lemma \ref{er_brig}}, \label{GerGb-ton-tac3} \\ \hspace{-5pt}\textstyle{\frac{1}{2}n(n-1)P_n {x_n}^2 \cdot [1 \pm o(1)]}\hspace{-.5pt},&\text{\hspace{-17pt}for Lemma \ref{er_brig-weaker}}. \label{GerGb-ton-tac4}
% \end{subnumcases}

Now based on (\ref{varYEY}) and (\ref{GerGb-ton-tac2}), we provide a lower
bound on $Y$ with high probability. By Chebyshev's inequality, \f
for any $\phi > 0$,
\begin{align}
\mathbb{P}\big[\hspace{1pt}|Y-\mathbb{E}[Y]|\geq
\phi\sqrt{\textrm{Var}[Y]}\hspace{1pt}\big] \leq {\phi}^{-2}.
\end{align}
We select% $\phi  =
%\frac{\left\{\mathbb{E}[Y]\right\}^{\frac{5}{6}}}{2\sqrt{\textrm{Var}[Y]}}$,
\begin{align}
 \phi & =
\frac{\big\{\mathbb{E}[Y]\big\}^{\frac{5}{6}}}{2\sqrt{\textrm{Var}[Y]}},
\label{phieq}
\end{align}
 which with (\ref{varYEY}) and (\ref{GerGb-ton-tac2}) results in $\phi  =
\omega(1)$ and hence
\begin{align}
\mathbb{P}\big[ \hspace{1pt}Y < \mathbb{E}[Y] -
\phi\sqrt{\textrm{Var}[Y]}\hspace{1pt}\big] = o(1). \label{Y}
\end{align}

Let $Z$ be a Poisson random variable with mean %${\lambda_n}$ defined by
\begin{align}
{\lambda_n} & : = \mathbb{E}[Y] -
\big\{\mathbb{E}[Y]\big\}^{\frac{5}{6}}. \label{lam}
\end{align}
With ${\psi_n}$ defined by
\begin{align}
 {\psi_n} & : =
\m{\frac{1}{2}} \big\{\mathbb{E}[Y]\big\}^{\frac{1}{3}}, \label{psi}
\end{align}
 we conclude from (\ref{GerGb-ton-tac2}) (\ref{lam}) and (\ref{psi}) that
${\psi_n} = \omega(1)$ and ${\psi_n} =
o\big(\sqrt{{\lambda_n}}\hspace{2pt}\big)$.

By \cite[Lemma 1.2]{Citeulike:505396}, \h\vspace{-2pt}
\begin{align}
 \mathbb{P}\big[\hspace{1pt}Z \geq {\lambda_n} + {\psi_n}
\sqrt{{\lambda_n}}\hspace{2pt}\big]  \leq e^{{\psi_n} \sqrt{{\lambda_n}}
-({\lambda_n} + {\psi_n} \sqrt{{\lambda_n}})
\ln(1+\frac{{\psi_n}}{\sqrt{{\lambda_n}}})}. \label{poisax}
\end{align}
From ${\psi_n} = o\big(\sqrt{{\lambda_n}}\hspace{2pt}\big)$, then for all
$n$ sufficiently large, we have
$\ln\big(1+\frac{{\psi_n}}{\sqrt{{\lambda_n}}}\big) \geq
\frac{{\psi_n}}{\sqrt{{\lambda_n}}}-\frac{{\psi_n}^2}{2{\lambda_n}}$ (derived from a
Taylor expansion), which is used in (\ref{poisax}) to
yield\vspace{-2pt}
\begin{align}
 \mathbb{P}\big[\hspace{1pt}Z \geq {\lambda_n} + {\psi_n}
\sqrt{{\lambda_n}}\hspace{2pt}\big]  & \leq e^{{\psi_n} \sqrt{{\lambda_n}}
-({\lambda_n} + {\psi_n} \sqrt{{\lambda_n}})
\big(\frac{{\psi_n}}{\sqrt{{\lambda_n}}}-\frac{{\psi_n}^2}{2{\lambda_n}}\big)}
\nonumber
\\ & =
e^{\frac{{\psi_n}^2}{2}\big(\frac{{\psi_n}}{\sqrt{{\lambda_n}}}-1\big)}.
\label{pois}
\end{align}
Applying ${\psi_n} = \omega(1)$ and ${\psi_n} =
o\big(\sqrt{{\lambda_n}}\hspace{2pt}\big)$ to (\ref{pois}), we
obtain\vspace{-2pt}
\begin{align}
 \mathbb{P}\big[\hspace{1pt}Z \geq {\lambda_n} + {\psi_n}
\sqrt{{\lambda_n}}\hspace{2pt}\big] &  = o(1). \label{Z}
\end{align}

From (\ref{phieq}) (\ref{lam}) and (\ref{psi}), we
establish\vspace{-2pt}
\begin{align}
 {\lambda_n} + {\psi_n} \sqrt{{\lambda_n}} & \leq \mathbb{E}[Y] -
\big\{\mathbb{E}[Y]\big\}^{\frac{5}{6}} + \m{\frac{1}{2}}
\big\{\mathbb{E}[Y]\big\}^{\frac{1}{3}} \cdot \sqrt{\mathbb{E}[Y]}
\nonumber
\\   & =  \mathbb{E}[Y] -
\phi\sqrt{\textrm{Var}[Y]}. \label{lambda}
\end{align}

Given (\ref{Y}) (\ref{Z}) and (\ref{lambda}), we obtain\vspace{-2pt}
\begin{align}
 &\mathbb{P}[ Y \geq Z] \nonumber
\\  & \geq \mathbb{P}\Big[
 \big(Y \geq \mathbb{E}[Y] - \phi\sqrt{\textrm{Var}[Y]}\hspace{1pt}
\big) \bcap  \hspace{1pt} ({\lambda_n} + {\psi_n} \sqrt{{\lambda_n}} \geq
Z\hspace{1pt} )\Big] \nonumber
\\ & \geq 1 - \mathbb{P}\big[Y < \mathbb{E}[Y] -
\phi\sqrt{\textrm{Var}[Y]}\hspace{1pt}\big] - \mathbb{P}\big[
{\lambda_n} + {\psi_n} \sqrt{{\lambda_n}} < Z\hspace{1pt}\big] \nonumber
\\ & \to 1,\textrm{ as }n \to \infty, \label{YZ}
\end{align}
where in the second to the last step, we use a union bound.

Given (\ref{YZ}), by the definition of graph $L_d(n,X)$, it is easy
to construct a coupling such that $L_d(n,Z)$ is a subgraph of
$L_d(n,Y)$ with probability $1-o(1)$; namely,
\begin{align}
 L_d(n,Y)& \succeq_{1-o(1)}L_d(n,Z) . \label{HqnZ}
\end{align}

From \cite[Proof of Claim 1]{Fill:2000:RIG:340808.340814}, for
Poisson random variable $Z$ with mean ${\lambda_n}$, in sampling $Z$
edges with repetition from all possible $\binom{n}{2}$ edges of an
$n$-size node set, the numbers of draws for different edges are
independent Poisson random variables with mean
\begin{align}
{\mu_n}:={\lambda_n}\Bigg/\binom{n}{2},\label{defmu}
\end{align}
where ``with repetition'' means that  at each time, an edge is selected from the $\binom{n}{2}$ edges, so we have that even if an edge has already been selected, it may get selected again next time.
 Therefore, $L_d(n,Z)$ with $Z \in \textrm{Poisson}({\lambda_n})$ is an
Erd\H{o}s--R\'{e}nyi graph \cite{citeulike:4012374} in which each
edge independently appears with a probability that a Poisson random
variable with mean ${\mu_n}$ is at least $d$, i.e., a probability of
\begin{align}
\varrho_n &  : =\sum_{x=d}^{\infty} \frac{{{\mu_n}}^x e^{-{\mu_n}}}{x!}.
\label{pndef}
\end{align}

In view that $L_d(n,Z)$ is equivalent to $G(n,\varrho_n)$, then
from (\ref{HqnY}) and (\ref{HqnZ}), \f
\begin{align}
 H_d(n,P_n,x_n)& \succeq_{1-o(1)}G(n,\varrho_n) , \label{gercp}
\end{align}
which is exactly (\ref{GerGb}) in Lemma
\ref{er_brig}. Therefore, to complete proving
Lemma \ref{er_brig}, we now analyze $\varrho_n$ in (\ref{pndef}).

% the proof of Lemma \ref{er_brig} is
% completed once we show that $\varrho_n$ defined in (\ref{pndef})
% satisfies (\ref{pnpb01}) (i.e., $\varrho_n   = p_b \cdot [1\pm
% o(1)]$).

From \cite[Proposition 1]{PES:36149}, $\varrho_n$ in (\ref{pndef})
can be bounded by\vspace{-1pt}
\begin{align}
\frac{{{\mu_n}}^d e^{-{\mu_n}}}{d!}  & < \varrho_n < \frac{{{\mu_n}}^d
e^{-{\mu_n}}}{d!} \cdot\bigg(1-\frac{{\mu_n}}{d+1}\bigg)^{-1}.\vspace{-1pt}
\label{pnbound}
\end{align}

To evaluate $\varrho_n$ based on (\ref{pnbound}), we now assess $\mu_n$ in (\ref{defmu}), and   analyze $\lambda_n$ in (\ref{lam}).
Applying (\ref{GerGb-ton-tac4}) and (\ref{GerGb-ton-tac2}) to (\ref{lam}), and noting that $\big[1 \pm o\big(\frac{1}{\ln n}\big)\big]\cdot \big[1 \pm o\big(\frac{1}{\ln n}\big)\big]$ (resp., \vspace{1pt} $[1\pm o(1)] \cdot  [1\pm o(1)]$) can also be written as $\big[1 \pm o\big(\frac{1}{\ln n}\big)\big]$ (resp., $[1\pm o(1)]$), we obtain
\begin{align}
 {\lambda_n} &    = \mathbb{E}[Y] -
\big\{\mathbb{E}[Y]\big\}^{\frac{5}{6}} \nonumber \\    & = \mathbb{E}[Y] \cdot \Big[1-
\big\{\mathbb{E}[Y]\big\}^{-\frac{1}{6}}\Big]\nonumber \\    & =\textstyle{\frac{1}{2}n(n-1)P_n {x_n}^2 \cdot \big[1 \pm o\big(\frac{1}{\ln n}\big)\big]}.\label{GerGb-ton-tac6}
\end{align}
% \begin{subnumcases}{\hspace{-11pt}    = }
% \hspace{-5pt}\textstyle{\frac{1}{2}n(n-1)P_n {x_n}^2 \cdot \big[1 \pm o\big(\frac{1}{\ln n}\big)\big]}\hspace{-.5pt},&\text{\hspace{-17pt}for Lemma \ref{er_brig}}, \label{GerGb-ton-tac5} \\ \hspace{-5pt}\textstyle{\frac{1}{2}n(n-1)P_n {x_n}^2 \cdot [1 \pm o(1)]}\hspace{-.5pt},&\text{\hspace{-17pt}for Lemma \ref{er_brig-weaker}}. \label{GerGb-ton-tac6}
% \end{subnumcases}
% \begin{subnumcases}{{\mu_n}  = }
% \hspace{-5pt}\textstyle{ P_n {x_n}^2 \cdot \big[1 \pm o\big(\frac{1}{\ln n}\big)\big]}\hspace{-.5pt},&\text{\hspace{-17pt}for Lemma \ref{er_brig}}, \label{GerGb-ton-tac7} \\ \hspace{-5pt}\textstyle{ P_n {x_n}^2 \cdot [1 \pm o(1)]}\hspace{-.5pt},&\text{\hspace{-17pt}for Lemma \ref{er_brig-weaker}}. \label{GerGb-ton-tac8}
% \end{subnumcases}
The application of (\ref{GerGb-ton-tac6}) to (\ref{defmu}) gives
\begin{align}
{\mu_n}  =
\textstyle{ P_n {x_n}^2 \cdot \big[1 \pm o\big(\frac{1}{\ln n}\big)\big]}.\label{GerGb-ton-tac8}
\end{align}
Note that Lemma \ref{er_brig} has condition (\ref{er_brig-eq3}) (i.e., ${x_n}^2 P_n = o\left( \frac{1}{\ln n} \right)$). Using this in (\ref{GerGb-ton-tac8}), we have
\begin{align}
{\mu_n}  =
\textstyle{o\left( \frac{1}{\ln n} \right)}.\label{GerGb-ton-tac10}
\end{align}

For any sequence $a_n$ satisfying ${a_n} = \pm o(1)$, we explain below $(1+a_n)^d = 1 \pm \Theta(a_n)$ since $d$ is a constant. To see this, given $|a_n|<1$ for all $n$ sufficiently large from ${a_n} = \pm o(1)$, we obtain: on the one hand, $(1+a_n)^d \leq (1+|a_n|)^d = 1 + \sum_{i=1}^d \big[ \binom{d}{i} |a_n|^i \big] \leq 1 + |a_n|\sum_{i=1}^d   \binom{d}{i} = 1 + (2^d-1) |a_n| =  1 + \Theta(a_n)$; on the other hand, $(1-a_n)^d \leq (1+|a_n|)^d = 1 + \sum_{i=1}^d \big[ \binom{d}{i} (-|a_n|)^i \big] \geq 1 - |a_n|\sum_{i=1}^d   \binom{d}{i} = 1 - (2^d-1) |a_n| =  1 - \Theta(a_n)$. Summarizing $ 1 - \Theta(a_n) \leq  (1-a_n)^d \leq  1 + \Theta(a_n)$, we obtain
 \begin{align}
(1+a_n)^d = 1 \pm \Theta(a_n)\text{ for }{a_n} = \pm o(1).  \label{pnbounda-ton-tac-tonasab}
\end{align}
From (\ref{GerGb-ton-tac8}) and (\ref{pnbounda-ton-tac-tonasab}), it holds  that
% \begin{subnumcases}{{\mu_n}^d  = }
% \hspace{-5pt}\textstyle{ (P_n {x_n}^2)^d \cdot \big[1 \pm o\big(\frac{1}{\ln n}\big)\big]}\hspace{-.5pt},&\text{\hspace{-17pt}for Lemma \ref{er_brig}}, \label{GerGb-ton-tac7-tif} \\ \hspace{-5pt}\textstyle{ (P_n {x_n}^2)^d \cdot [1 \pm o(1)]}\hspace{-.5pt},&\text{\hspace{-17pt}for Lemma \ref{er_brig-weaker}}. \label{GerGb-ton-tac8-tif}
% \end{subnumcases}
\begin{align}
{\mu_n}^d  =
\textstyle{ (P_n {x_n}^2)^d \cdot \big[1 \pm o\big(\frac{1}{\ln n}\big)\big]}.\label{GerGb-ton-tac8-tif}
\end{align}

For ${\mu_n} = o(1)$, we explain below $e^{-{\mu_n}} = 1 - \Theta({\mu_n})$. To see this, on the one hand, it holds that $e^{-{\mu_n}} \geq 1 - {\mu_n}$. On the other hand, given ${\mu_n} < 0.5$ for all $n$ sufficiently large (which holds from ${\mu_n} = o(1)$), we can easily show $e^{-{\mu_n}} \leq 1 - 0.5 {\mu_n}$ by taking the derivative of $e^{-{\mu_n}} - (1 - 0.5 {\mu_n})$ to investigate its monotonicity. Summarizing $ 1 - {\mu_n} \leq  e^{-{\mu_n}} \leq 1 - 0.5 {\mu_n}$, we obtain
 \begin{align}
e^{-{\mu_n}} = 1 - \Theta({\mu_n}).  \label{pnbounda-ton-tac-tona}
\end{align}

From ${\mu_n} = o(1)$, we have $\big(1-\frac{{\mu_n}}{d+1}\big)^{-1} = 1 + \frac{{\mu_n}}{d+1-{\mu_n}} = 1 + \Theta({\mu_n})$, which along with (\ref{pnbounda-ton-tac-tona}) is used in (\ref{pnbound}) to derive
\begin{align}
\varrho_n & \hspace{-1pt}=\hspace{-1pt} \frac{{{\mu_n}}^d e^{-{\mu_n}}}{d!} \hspace{-1pt} \cdot \hspace{-1pt}\big[1\hspace{-1pt} +\hspace{-1pt} \Theta({\mu_n})\big]\hspace{-1pt} = \hspace{-1pt} \frac{{{\mu_n}}^d}{d!}  \hspace{-1pt}\cdot \hspace{-1pt}\big[1\hspace{-1pt} - \hspace{-1pt}\Theta({\mu_n})\big]\hspace{-1pt}  \cdot\hspace{-1pt}\big[1\hspace{-1pt} + \hspace{-1pt}\Theta({\mu_n})\big].  \label{pnbounda-ton-tac-tonabc}
\end{align}
For any two sequences $c_n$ and $d_n$ satisfying
$c_n = \Theta({\mu_n})$ and $d_n = \Theta({\mu_n})$ with ${\mu_n} = o(1)$, we have $(1-c_n) (1+d_n) = 1 - c_n + d_n - c_n d_n = 1 \pm \Theta({\mu_n})$, which we use in (\ref{pnbounda-ton-tac-tonabc}) to get
\begin{align}
\varrho_n & = \frac{{{\mu_n}}^d}{d!}  \cdot \big[1 \pm  \Theta({\mu_n})\big] .  \label{pnbounda-ton-tac-tonabc5}
\end{align}
Then applying (\ref{GerGb-ton-tac8-tif}) and (\ref{GerGb-ton-tac10}) to (\ref{pnbounda-ton-tac-tonabc5}), and noting that $\big[1 \pm o\big(\frac{1}{\ln n}\big)\big]\cdot \big[1 \pm o\big(\frac{1}{\ln n}\big)\big]$ (resp., \vspace{1pt} $[1\pm o(1)] \cdot  [1\pm o(1)]$) can also be written as $\big[1 \pm o\big(\frac{1}{\ln n}\big)\big]$ (resp., $[1\pm o(1)]$), we obtain
\begin{align}
\varrho_n & = \textstyle{\frac{(P_n{x_n}^2)^d}{d!}  \cdot \big[1 \pm o\big(\frac{1}{\ln n}\big)\big]} . \label{GerGb-ton-tac11}
\end{align}

% \begin{align}
% \varrho_n & = \frac{{{\mu_n}}^d}{d!}  \cdot \big[1 \pm o\big(\frac{1}{\ln n}\big)\big]=
% \frac{(P_n{x_n}^2)^d}{d!}  \cdot \big[1 \pm o\big(\frac{1}{\ln n}\big)\big].  \label{pnbounda-ton-tac-ton}
% \end{align}

% ==========

% From (\ref{varYEY2}) (\ref{lam}) (\ref{defmu}), and conditions
% $n^2{x_n}^2 P_n
%  = \omega(1)$ and $P_n{x_n}^2 = o(1)$, \f \vspace{-1pt}
% \begin{align}
% {\mu_n}:&={\lambda_n}\Bigg/\binom{n}{2} = P_n{x_n}^2
% \Big[1-O\Big((n^2{x_n}^2 P_n)^{-\frac{1}{6}}\Big)\Big] \nonumber \\
% & = P_n{x_n}^2\cdot[1-o(1)] = o(1) . \label{mueq}
% \end{align}

% Using (\ref{mueq}) in (\ref{pnbound}), we obtain\vspace{-1pt}
% \begin{align}
% \varrho_n & \sim \frac{{{\mu_n}}^d e^{-{\mu_n}}}{d!} \sim
% \frac{(P_n{x_n}^2)^d}{d!} .  \label{pnbounda}
% \end{align}

From \cite[Fact 3]{zz}, for Erd\H{o}s--R\'enyi graphs
$G(n,y_n')$ and $G(n,y_n'')$, if $y_n ' \geq y_n''$, then
$G(n,y_n')\succeq G(n,y_n'')$. Therefore, by (\ref{gercp})
(\ref{GerGb-ton-tac11}) and \cite[Fact 3]{2013arXiv1301.0466R} on the
transitivity of graph coupling, we can set $y_n   =
\textstyle{\frac{(P_n{x_n}^2)^d}{d!}} \cdot \big[1 - o\big(\frac{1}{\ln n}\big)\big]$ to have
$ H_d(n,P_n,x_n) \succeq_{1-o(1)}G(n,y_n) $, so  that Lemma \ref{er_brig} is finally proved. \qeda

\end{document}